\documentclass{sig-alternate-05-2015}
\usepackage{amsmath,amsfonts,amssymb}
\usepackage{color}
\usepackage{graphicx}
\usepackage{epsfig}
\usepackage{url}
\usepackage{semantic}
\usepackage{subfigure}
\newcommand{\figref}[1]{Figure \ref{#1}}
\newcommand{\tabref}[1]{Table \ref{#1}}

\newcommand{\KZ}[1]{\textcolor{blue}{(KZ: #1)}}

\newcommand{\cut}[1]{}
\newcommand{\dd}[0]{\mathrm{d}}
\newcommand{\numberthis}{\stepcounter{equation}\tag{\theequation}}

\usepackage{courier}
\usepackage{listings}
\usepackage{etoolbox}
\makeatletter
\patchcmd{\maketitle}{\@copyrightspace}{}{}{}
\makeatother

\mathlig{=>}{\Rightarrow}
\mathlig{->}{\rightarrow}

\newcounter{enum}
\newenvironment{packed_enum}{
\begin{list}{(\alph{enum})}{
  \setlength{\itemsep}{-0.5pt}
  \setlength{\parskip}{1pt}
  \setlength{\labelwidth}{30 pt}
  \setlength{\leftmargin}{15 pt}
  \setlength{\itemindent}{0pt}
  \usecounter{enum}}
}{\end{list}}

\makeatletter
\newcommand{\Spvek}[2][r]{%
  \gdef\@VORNE{1}
  \left(\hskip-\arraycolsep%
    \begin{array}{#1}\vekSp@lten{#2}\end{array}%
  \hskip-\arraycolsep\right)}

\def\vekSp@lten#1{\xvekSp@lten#1;vekL@stLine;}
\def\vekL@stLine{vekL@stLine}
\def\xvekSp@lten#1;{\def\temp{#1}%
  \ifx\temp\vekL@stLine
  \else
    \ifnum\@VORNE=1\gdef\@VORNE{0}
    \else\@arraycr\fi%
    #1%
    \expandafter\xvekSp@lten
  \fi}
\makeatother

\newcommand\Mark[1]{\textsuperscript{#1}}

\begin{document}

\title{InferSpark: Statistical Inference at Scale}

%\numberofauthors{5}
\author{
Zhuoyue Zhao~\Mark{1}, Jialing Pei~\Mark{1}, Eric Lo~\Mark{2}, Kenny Q. Zhu~\Mark{1}, Chris Liu~\Mark{2}\\
\affaddr{\Mark{1}Shanghai Jiao Tong University \hspace*{4mm}\Mark{2}Hong Kong Polytechnic University}\\
\email{
\{zzy7896321@, peijialing@, kzhu@cs\}.sjtu.edu.cn
\hspace*{2mm}\{ericlo, cscyliu\}@comp.polyu.edu.hk
}
}

\maketitle

%\unitlength1pt
%\begin{picture}(0,0)
%\put(380,190){\mbox{\Large \bf  Paper \#478}}
%\end{picture}
%\unitlength1cm

\begin{abstract}
%\KZ{
The Apache Spark stack has enabled fast
large-scale data processing.
Despite a rich library of statistical models and
inference algorithms, it does not give domain
users the ability to develop their own models.
%}
The emergence of probabilistic programming languages 
has showed the promise of developing sophisticated
probabilistic models in a succinct and programmatic way.
%helps data analysts and machine learning experts to concisely 
%describe the probabilistic models using a programming language. 
These frameworks have the potential of automatically generating
inference algorithms for the user defined models and 
answering various statistical queries about the model. 
It is a perfect time to unite these two great directions to
produce a programmable big data analysis framework. 
We thus propose, InferSpark, a probabilistic programming framework on top of Apache Spark. 
Efficient statistical inference can be easily implemented on this 
framework and inference process can leverage the distributed main memory processing 
power of Spark. This framework makes statistical inference on
big data possible and speed up the penetration of probabilistic 
programming into the data engineering domain. 
\end{abstract}

%!TEX root = paper.tex

\section{Introduction}
\label{sec:intro}

Statistical inference is an important technique to express hypothesis and
reason about data in data analytical tasks.  
Today, many big data applications are based on
statistical inference.
Examples include topic modeling \cite{blei2003latent,Titov2008a},
sentiment analysis \cite{Titov2008b, Jo2011,tsm}, spam filtering \cite{spam}, to name a few.

One of most critical steps of statistical inference is to construct a
\emph{statistical model} to formally represent the underlying statistical
inference task \cite{cox}.  The development of a statistical
model is never trivial because a domain user may have to devise and
implement many  different models before finding a promising one for a specific
task.  Currently, most scalable machine learning libraries (e.g. MLlib \cite{mllib}) only
contain standard models like support vector machine, linear regression, latent
Dirichlet allocation (LDA) \cite{blei2003latent}, etc.  
To carry out statistical inference on
customized models with big data, the user has to implement her own models and
inference codes on a distributed framework like Apache Spark
\cite{Zaharia:2010:SCC:1863103.1863113} and Hadoop \cite{hadoop}.

Developing inference code requires extensive knowledge in both statistical
inference and programming techniques in distributed frameworks.  Moreover,
model definitions, inference algorithms, and data processing tasks are all
mixed up in the resulting code, making it hard to debug and reason about.  For
even a slight alteration to the model in quest of the most promising one, the
model designer will have to re-derive the formulas and re-implement the
inference codes, which is tedious and error-prone. 

In this paper, we present InferSpark, a \emph{probabilistic programming
framework} on top of Spark.  Probabilistic programming is an emerging
paradigm that allows statistician and domain users to succinctly express a model
definition within a host programming language and transfers the burden of
implementing the inference algorithm from the user to the compilers and
runtime systems \cite{pp}.  For example, Infer.NET \cite{InferNET14} is a
probabilistic programming framework that extends C\#.  The user can express,
say, a Bayesian network in C\# and the compiler will generate code to perform
inference on it. Such code could be as efficient as the implementation of
the same inference algorithm carefully optimized by
an experienced programmer.

So far, the emphasis of probabilistic programming has been put on the
expressiveness of the languages and the development of efficient inference
algorithms (e.g., variational message passing \cite{vmp}, Gibbs sampling \cite{gibbs},
Metropolis-Hastings sampling \cite{mh}) to handle a wider range of statistical
models.  The issue of scaling out these frameworks, however, has not been
addressed.  For example, Infer.NET only works on a single machine.  
When we tried to use Infer.NET to train an LDA model of 96 topics and 9040-word
vocabulary on only 3\% of Wikipedia articles, the actual memory
requirement has already exceeded 512GB, the maximum memory of most commodity
servers today.
%Frequent swapping makes each iteration take xxx hrs
%(still running, > 2 hr). Infer .NET deals with the scaling problem by
%splitting the whole dataset into chunks that fit in memory and iteratively
%load and process the chunks. Using the batched training, each iteration still
%takes over 21 minutes.  
The goal of InferSpark is thus to bring 
probabilistic programming to Spark, a predominant distributed data
analytic platform, for carrying out statistical inference at scale. 
The InferSpark project consists of two parts:

\begin{packed_enum}
	\item {\bf Extending Scala to support probabilistic programming}

	Spark is implemented in Scala due to its functional nature.
The fact that both preprocessing and post-processing can be 
included in one Scala program substantially eases the development process.
	In InferSpark,
	we extend Scala with probabilistic programming constructs to leverage its
	functional features.  Carrying out statistical inference with InferSpark
	is simple and intuitive, and implicitly enjoys the distributed computing
	capability brought by Spark.  As an example, the LDA statistical model was
	implemented using 503 lines of Scala code in MLlib (excluding comments,
	javadocs, blank lines, and utilities of MLlib).  With InferSpark, we could
	implement that using only 7 lines of Scala code (see
	\figref{fig:intro_lda_def}).

	\item {\bf Building an InferSpark compiler and a runtime system}
		
	InferSpark compiles InferSpark models into Scala classes
	and objects that implement the inference algorithms with a set of API. The
	user can call the API from their Scala programs to specify the input
	(observed) data and query about the model (e.g. compute the expectation of
	some random variables or retrieve the parameters of the posterior
	distributions).
		
\end{packed_enum}

\begin{figure}
\begin{lstlisting}
@Model
class LDA(K: Long, V: Long, alpha: Double, beta: Double){
	val phi = (0L until K).map{_ => Dirichlet(beta, K)}
	val theta = ?.map{_ => Dirichlet(alpha, K)}
	val z = theta.map{theta => ?.map{_ => Categorical(theta)}}
	val x = z.map{_.map{z => Categorical(phi(z))}}
}
\end{lstlisting}
\label{fig:intro_lda_def}
\caption{Definition of Latent Dirichlet Allocation Model}
\end{figure}

Currently, InferSpark supports Bayesian network models. Bayesian network
is a major branch of probabilistic graphical model and it has already covered
models like naive Bayes, LDA, TSM \cite{tsm}, etc.  The goal of this paper is to
describe the workflow, architecture, and Bayesian network implementation of
InferSpark.  We will open-source InferSpark and support other models (e.g.,
Markov networks) afterwards.  

To the best of our knowledge, InferSpark is the first endeavor to bring 
probabilistic programming into the (big) data engineering domain.
Efforts like MLI \cite{mli} and SystemML \cite{systemml} all aim 
at easing the difficulty of developing \emph{distributed machine learning techniques} 
(e.g., stochastic gradient descent (SGD)).
InferSpark aims at easing the complexity of developing \emph{custom statistical models}, 
with statistician, data scientists, and machine learning researchers as the target users.
This paper presents the following technical contributions of InferSpark so far.
\begin{packed_enum}
\item We present the extension of Scala's syntax that can express various sophisticated 
Bayesian network models with ease.
\item We present the details of compiling and executing an InferSpark program on Spark.
That includes the mechanism of automatic generating efficient inference codes that 
include checkpointing (to avoid long lineage), proper timing of caching and 
anti-caching (to improve efficiency under memory constraint),
 and partitioning (to avoid unnecessary replication and shuffling).
\item We present an empirical study that shows InferSpark can enable 
statistical inference on both customized and standard models at scale.
\end{packed_enum}

%
%For example, Bayesian inference algorithms are mostly iterative serial
%algorithms. When implementing them on Spark, we have to identify the
%parallelizable parts without violating the correctness of the algorithms and
%also adapt them to the computation model.
%% (\ZY{not the problem of
%%functional style. when updating a vertex $v$, $v$ needs to pull message from
%%neighbours $u$, and before the neighbours can return the message, they may
%%need to pull messages from their neighbours.
%%This translates to either 2 ``aggregate'' and 2
%%``join'', or 1 and 1 with cached message in GraphX.}). 
%Furthermore, 
%Spark is efficient for iterative machine learning algorithms 
%that does not modify the in-memory RDD (e.g., Stochastic Gradient Descent in MLlib). 
%%
%%because the data can reside in the memory as RDD,
%%
%%as long as the
%%data reside in the memory, especially for algorithms 
%Statistical inference algorithms like VMP \cite{vmp}, however, 
%require iteratively update the RDD.
%Therefore, the implementation requires persisting  the 
%intermediate results before the first action of the next
%iteration to avoid unnecessary shuffling and 
%unpersisting afterwards so that
%the memory consumption does not increase linearly with the number of iterations.
%It is also necessary to periodically checkpoint the RDD to avoid long lineage,
%which could overflow the memory and crash the program. \KZ{One might ask,
%is Spark the right choice to implement PP on, if there's so much difference
%between the intrinsic computation model of Spark and VMP?}

The remainder of this paper is organized as follows: 
Section \ref{sec:background} presents the essential background for this paper.
Section \ref{sec:framework} then gives an overview of InferSpark.
Section \ref{sec:implementation} gives the implementation details of InferSpark.
Section \ref{sec:eval} presents an evaluation study of the current version of InferSpark.
Section \ref{sec:related} discusses related work 
and Section \ref{sec:conclusion} contains our concluding remarks.

\section{Background}
\label{sec:background}
This section presents some preliminary knowledge about
statistical inference and, in particular, Bayesian inference using 
variational message passing, a popular variational inference algorithm, 
as well as its implementation concerns on Apache Spark/GraphX stack.
  
\subsection{Statistical Inference}

Statistical inference is a common machine learning task of obtaining the
properties of the underlying distribution of data. For example, one can infer
from many coin tosses the probability of the coin turning up head by counting
how many tosses out of the all tosses are head. There are two different
approaches to model the number of heads: the frequentist approach and the
Bayesian approach.

Let $N$ be the total number of tosses and $H$ be the number of heads.  In
frequentist approach, the probability of coin turning up head is viewed as an
unknown \emph{fixed} parameter so the best guess $\phi$ would be the number of heads $H$ in the
results over the total number of tosses $N$.
\begin{equation*}
	\phi = \frac{H}{N}
\end{equation*}

In Bayesian approach, the probability of head is viewed as a hidden \emph{random
variable} drawn from a prior distribution, e.g., $\mathrm{Beta}(1, 1)$, the
uniform distribution over [0, 1]. According to the Bayes Theorem, the
posterior distribution of the probability of coin turning up head can be
calculated as follows:

\begin{align*}
	p(\phi|x) &= \frac{ \phi^H (1-\phi)^{N-H} f(\phi; 1, 1)}{\int_0^1 \phi^H
	(1-\phi)^{N-H} f(\phi; 1, 1)\mathrm{d}\phi} \\ &= f(\phi; H+1, N-H+1) \numberthis
	\label{eqn:coin_posterior}
\end{align*}

\noindent
where $f(\cdot; \alpha, \beta)$ is the probability density function (PDF) of
$\mathrm{Beta}(\alpha, \beta)$ and $x$ is the outcome of $N$ coin tosses.

The frequentist approach needs smoothing and regularization techniques to
generalize on unseen data while the Bayesian approach does not because the
latter can capture the uncertainty by modeling the parameters as random
variables. 

\subsection{Probabilistic Graphical Model}

Probabilistic graphical model \cite{pgm} (PGM) is a graphical representation of the
conditional dependencies in statistical inference. Two types of PGM are widely
used: Bayesian networks and Markov networks. Markov networks are undirected
graphs while Bayesian networks are directed acyclic graphs. Each type of PGM
can represent certain independence constraints that the other cannot represent. 
InferSpark currently supports Bayesian networks and regards Markov networks 
as the next step.  
\begin{figure}[h]
	\centering
	\includegraphics[scale=0.38]{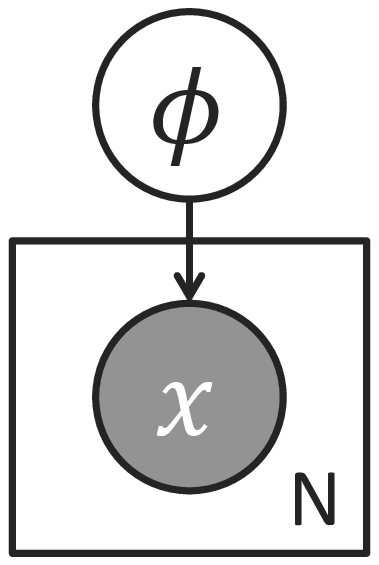}
	\caption{Bayesian network of the coin flip model (observed/unobserved random
	variable are in dark/white)}
	\label{fig:coin_bn}
\end{figure}

In a Bayesian network, the vertices are random variables and the edges
represent the conditional dependencies between the random variables.
The joint probability of a Bayesian network can be factorized into conditional
probabilities of each vertex $\theta$ conditioned on their parents
$\mathrm{F}(\theta)$. 
\figref{fig:coin_bn} is the Bayesian network of the coin flip model.  
Here, the factors in the joint
probability are $p(\phi)$ and $p(x|\phi)$. The plate surrounding $x$ represents
repetition of the random variables. The subscript $N$ is the number of
repetitions. The outcome of coin tosses $x$ is repeated $N$ times and
each depends on the probability $\phi$. The Bayesian network of the coin flip
model encodes the joint probability $p(\phi, x) =
p(\phi)\prod_{i=1}^{N}p(x_i|\phi)$. 

Bayesian networks are generative models, which describes the process of
generating random data from hidden random variables. The typical inference
task on generative model is to calculate the posterior distribution of the
hidden variables given the observed data. In the coin flip model, the
observed data are the outcomes of coin tosses and the hidden random variable
is the probability of head. The inference task is to calculate the posterior
in Equation \ref{eqn:coin_posterior}.

\subsection{Bayesian Inference Algorithms}

\begin{figure}
	\centering
	\includegraphics[scale=0.4]{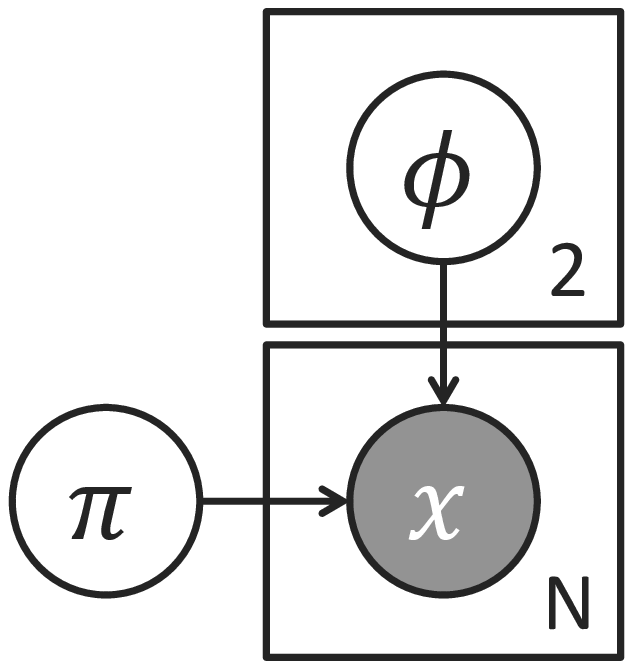}
	\caption{Bayesian network of the two-coin model}
	\label{fig:two_coins}
\end{figure}

Inference of the coin flip model is simple because the posterior
(Equation \ref{eqn:coin_posterior})
 has a tractable analytical solution. 
However, most real-world models are more complex than that
and their posteriors do not have a familiar form. 
Moreover, even calculating the
probability mass function or probability density function at one point is hard
because of the difficulty of calculating the probability of the
observed data in the denominator of the posterior. The probability of the
observed data is also called evidence. It is the summation or integration over
the space of hidden variables and is hard to calculate because of exponential
growth of the number of terms. 

Consider a two-coin model in \figref{fig:two_coins}, where we first decide
which coin to toss, with probability $\pi_1$ to choose the first coin and
probability $\pi_2$ to choose the second coin ($\pi_1 = 1 - \pi_2$). 
We then toss the chosen coin, which has probability $\phi_i$ to turn up head. 
This process is repeated $N$ times. The two-coin model is
a {\em mixture model}, which represents the
mixture of multiple sub-populations. Each such sub-population, in this case
$\phi_1$ and $\phi_2$, have their own distributions, 
while the observation can only be obtained on the overall population, that is
the number of heads after $N$ tosses. 
The two-coin model has no tractable analytical solution. 
Assuming Beta priors for $\pi$, $\phi_{1}$ and $\phi_{2}$, 
the posterior distribution is:

{
\small
\begin{align*}
%	&p(\pi, \phi, x) = \\
%	&f(\pi)f(\phi_1)f(\phi_2)(\pi_1 \phi_1 + \pi_2 \phi_2)^H (\pi_1
%	(1-\phi_1) + \pi_2 (1- \phi_2))^{N-H} \\
	&p(\pi, \phi | x) = \frac{p(\pi, \phi, x)}{\int p(\pi, \phi, x) \dd \pi \dd
\phi_1 \dd \phi_2}
\end{align*}
}

\noindent
where the joint distribution $p(\pi, \phi, x)$ is:

{\small
\begin{align*}
%	&p(\pi, \phi, x) = \\
	&f(\pi)f(\phi_1)f(\phi_2)(\pi_1 \phi_1 + \pi_2 \phi_2)^H (\pi_1
	(1-\phi_1) + \pi_2 (1- \phi_2))^{N-H} \\
\end{align*}
}

The integral in the denominator of the posterior is intractable 
because it has $2^N$ terms and takes exponential time to compute.  
Since solving for the exact posterior is
intractable, approximate inference algorithms are used instead.
Although approximate inference is also NP-hard, it performs well in 
practical applications. 
Approximate inference techniques include Markov Chain Monte Carlo (MCMC) 
method, variational inference and so on.
%MCMC algorithms are more
%accurate than the variational inference but the running time is usually much
%longer. 
MCMC algorithms are inherently non-deterministic, and single random number
generator is required to ensure randomness. In a distributed setting, sharing
a single random number generator across the nodes in a cluster is a serious
performance bottleneck. Having different generators on different nodes would
risk the correctness of the MCMC algorithms.
On the other hand, variational inference methods such as
Variational Message Passing (VMP) \cite{vmp}
%of exponential-conjugate Bayesian networks because it
%can be expressed as a 
is a deterministic graph-based message passing algorithm, 
which can be easily adapted to a distributed graph computation model such as
GraphX \cite{graphX}. 
InferSpark currently supports VMP.
Support of other techniques (e.g., MCMC) is included in InferSpark's open-source agenda.

\begin{figure}[ht]
	\centering
	\includegraphics[scale=0.4]{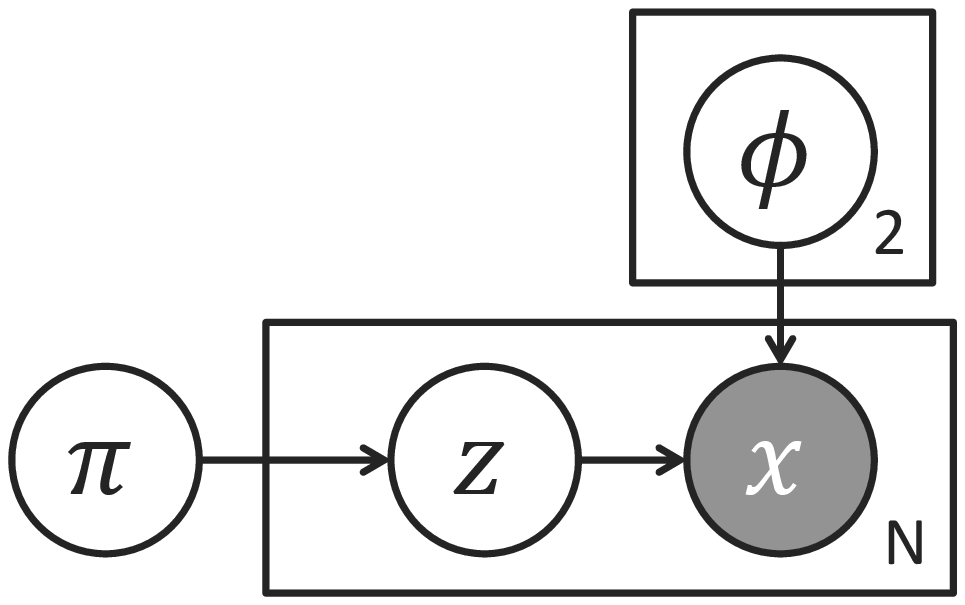}
	\caption{Expanded Bayesian network of the two-coin model}
	\label{fig:two_coin_bn}
\end{figure}

To infer the posterior of the two-coin model using VMP,
the original Bayesian network has to be expanded by 
adding some more hidden random variables.
Figure \ref{fig:two_coin_bn}
shows Bayesian network with hidden random variables added,
%illustrate the algorithm, we use the two-coin model
%(\figref{fig:two_coin_bn}), a slightly more complex model than the one-coin
%model. 
where $z_i$ is the index (1 or 2) of the coin chosen for the $i^{th}$ toss.

%$z_{ij}$ is an indicator random variable 
%where if the $j$-th coin is chosen
%in the $i$-th experiment,   then $z_{ij} = 1$, else $z_{ij} = 0$ otherwise.
%\ERIC{am I correct here, ZY?}

%
%Let $k \in \{1, 2\}$ be the indices of the coins. Coin $k$'s probability of turning head is $\phi_k$. 
%We repeat the experiment for $N=2$ times. 
%In the $i^{th}$ experiment, we first choose
%one of the coin with probabilities $\pi_1$ and $\pi_2 = 1 - \pi_1$. 
%Let the index of the chosen coin be $z_i$. 
%Then we flip the coin $z_i$ once and get a
%head or a tail, denoted as $x_i$. For any of the variables $v$ above, we use
%$v_j, j \in \dom(v)$ as an indicator variable for $v = j$. For example, $z_{i1}
%= 1$ if $z_i = 1$ and $z_{i1} = 0$ otherwise.

\begin{figure}[ht]
	\centering
	\includegraphics[scale=0.4]{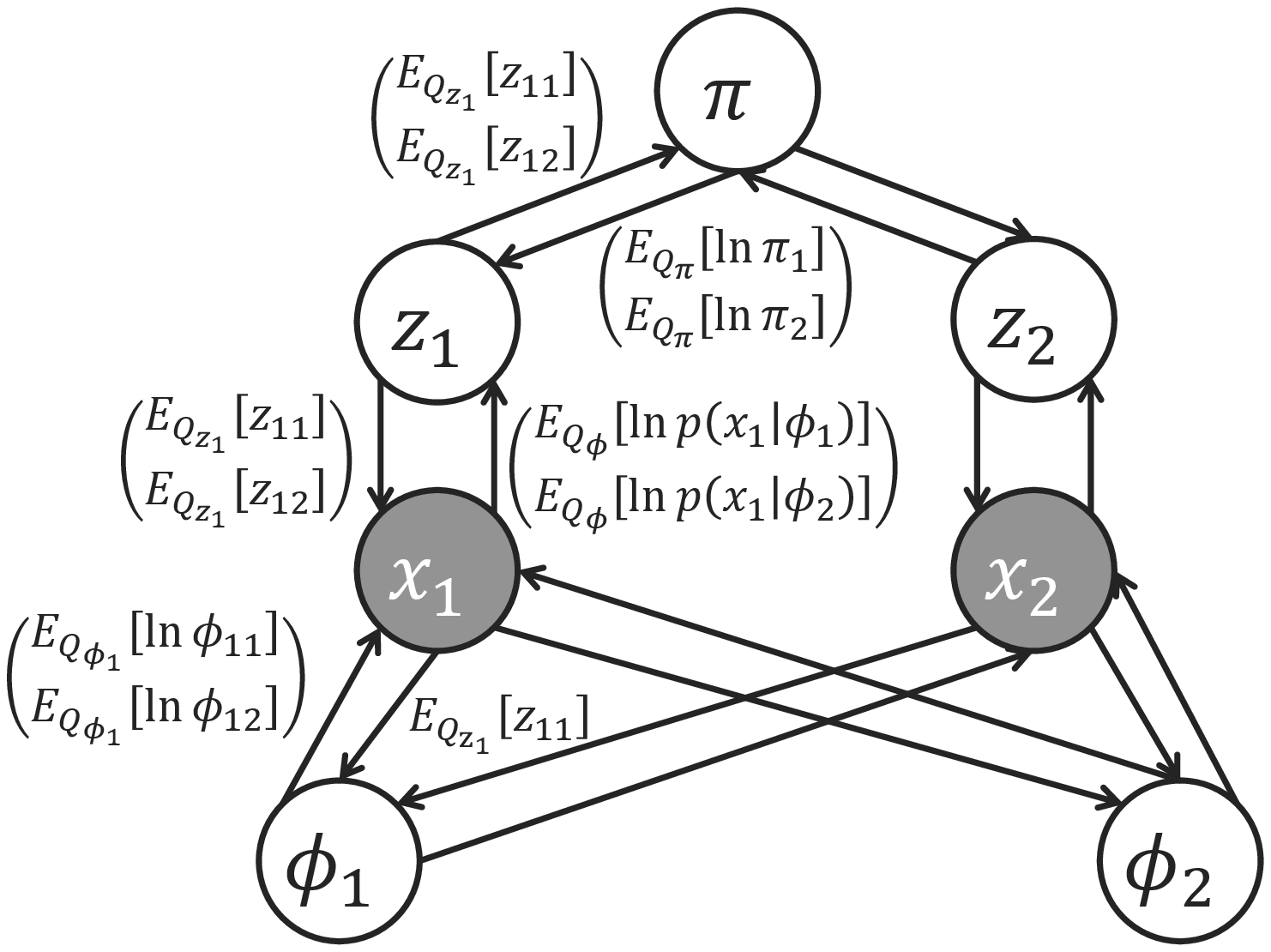}
	\caption{Message passing graph of the two-coin model}
	\label{fig:two_coins_mpg}
\end{figure}

The VMP algorithm approximates the posterior distribution with a fully
factorized distribution $Q$. 
The algorithm iteratively passes messages along
the edges and updates the parameters of each vertex to minimize the KL divergence
between $Q$ and the posterior.  
Because the true posterior is unknown, VMP algorithm maximizes the
evidence lower bound (ELBO), which is equivalent to minimizing the KL
divergence~\cite{vmp}. However, ELBO involves only the expectation 
of the log likelihoods of the approximated distribution $Q$ and 
is thus straightforward to compute. 

\figref{fig:two_coins_mpg} shows the \emph{message passing graph} of
VMP for the two-coin model with $N=2$ tosses.
There are four types of vertices in the two-coin model's message passing graph: $\phi$, $\pi$, $z$, and $x$, 
with each type corresponding to a variable in the Bayesian network.
Vertices in the Bayesian network are expanded in the message passing graph.
For example, the repetition of vertex $\phi$ is 2 in the model.
So, we have $\phi_1$ and $\phi_2$ in the massage passing graph.
Edges in the two-coin model's message passing graph are bidirectional
and different messages are sent in different directions.
The three edges $\pi \rightarrow z$, 
$z \rightarrow x$, $\phi_k \rightarrow x$ in the Bayesian network
thus create six types of edges in the message passing graph:
$\pi \rightarrow z_i$, 
$\pi \leftarrow z_i$, 
$z_i \rightarrow x_i$, 
$z_i \leftarrow x_i$, 
$\phi_k \rightarrow x_i$, and
$\phi_k \leftarrow x_i$.

%all the random variables in the two-coin model with messages of the 
%algorithm annotated to the edges.  There are 4 types of vertices and 6 types
%of messages in the VMP algorithm for two-coin model, for each of which we only
%show one instance. 

Each variable (vertex) is associated with the parameters of its
approximate distribution. Initially the parameters can be arbitrarily
initialized. For edges whose direction is the same as in the Bayesian network,
%(i.e. $\pi \rightarrow z_i$, $z_i \rightarrow x_i$, $\phi_k \rightarrow x_i$),
the message content only depends on the parameters of the sender.  
For example, the
message $m_{\pi \rightarrow z_1}$ from $\pi$ to $z_1$ is a vector of expectations of logarithms of
$\pi_1$ and $\pi_2$,
denoted as $\Spvek{E_{Q_\pi}[\ln \pi_1];E_{Q_\pi}[\ln \pi_2]}$ in \figref{fig:two_coins_mpg}.
For edges whose direction is opposite of those in the
Bayesian network, %(e.g., $z_i \rightarrow \pi$) %, $x_i \rightarrow z_i$, $x_i \rightarrow \phi_k$), 
in addition to the parameters of the sender, 
the message content may also depend on other parents of the sender in the Bayesian network. 
For example, the message $m_{x_1 \rightarrow z_1}$ from $x_1$ to $z_1$ is
$\Spvek{E_{Q_\pi}[\ln p(x_1 | \phi_1)];E_{Q_\pi}[\ln p(x_1 | \phi_2)]}$,
which depends both on the observed outcome $x_1$ and the
expectations of $\ln \phi_1$ and $\ln \phi_2$.  

Based on the message passing graph,
VMP selects one vertex $v$ in each iteration and 
pulls messages from $v$'s neighbor(s).
If the message source $v_s$ is the child of $v$ in the Bayesian network,
VMP also pulls message from $v_s$'s other parents. 
For example, assuming VMP selects $z_1$ in an iteration,
it will pull messages from $\pi$ and $x_1$.
Since $x$ is the child of $z$ in the Bayesian network (\figref{fig:two_coin_bn}), 
and $z$ depends on $\phi$,
VMP will first pull a message from $\phi_1$ to $x_1$,
then pull a message from $x_1$ to $z_1$.
This process however is would not be propagated and is restricted to 
only $v_s$'s direct parents.
%However, no more than
%2 rounds of messages need to be sent because message from parent to child only
%depends on the parent. 
On receiving all the requested messages, 
the selected vertex updates its parameters by aggregating the messages. 
%It has been proven that each iteration will reduce the KL divergence between the approximate
%distribution $Q$ and the posterior distribution. 

Implementing the VMP inference code for a statistical model 
(Bayesian network) $M$
requires (i) deriving all the messages mathematically  (e.g., deriving $m_{x_1 \rightarrow z_1}$)
and (ii) coding the message passing mechanism specifically for $M$.
%Implementing the VMP algorithm for an arbitrary exponential-conjugate Bayesian
%network involves deriving all the messages and updates and tranlating the
%mathematics into a real program. 
The program tends to contain a lot of boiler-plate code 
because there are many types of messages and vertices.
For the two-coin model, 
$x$ would be coded as a vector of integers 
whereas $z$ would be coded as a vector of probabilities (float), etc.
Therefore, even a slight alteration to the model, say, from two-coin to 
two-coin-and-one-dice, all the messages have to be re-derived 
and the message passing code has to be re-implemented, which is tedious
to hand code and hard to maintain.

%They
%can be vectors or scalars, floating point numbers or integers. For each
%individual case, specific code needs to be written. A few lines of
%mathematical derivations could easily grow into tens or even hundreds lines of
%code.
%

%To implement the VMP algorithm for an arbitrary exponential-conjugate model,
%the user first need to derive the ELBO $\mathcal{L}$  for the model. For each
%hidden random variable, separating out the terms int $\mathcal{L}$ relating to
%it to calculate the messages to pull from the neighbours.  Secondly,
%implementing the algorithm involves writing a lot of boiler-plate code. For
%each random variable, the user needs to write code to pull messages from all
%its neighbours. Several lines of mathematical expressions could expand into
%tens or even hundreds of lines of code of function definitions, swith-cases
%and loops. For the two-coins model, first step is to write down the ELBO
%(\eqnref{eqn:ELBO}) and derive the messages in \figref{fig:two_coins_mpg} for each
%edge according to (\eqnref{eqn:ELBO_one_term}). To translate the mathematical
%expressions into program, functions and data structures are declared. The user
%also has to write code to initialize each random variables. Finally, write a
%long loop with switch-cases to implement the iterative updates.

\subsection{Inference on Apache Spark}
When a domain user has crafted a new model $M$
and intends to program 
the corresponding VMP inference code on Apache Spark,
one natural choice is to do that through GraphX, the distributed graph processing framework on top of Spark.
Nevertheless, the user still has to go through a number of programming and system concerns,
which we believe, should better be handled by a framework like InferSpark instead.

First, %VMP-based inference code may require a vertex sends messages to its neighbors \emph{selectively}.  
the Pregel programming abstraction of GraphX
restricts that only updated vertices in the last iteration can send message in
the current iteration.
So for VMP,  
% but we can not always have all vertices that need to
%send messages active. For example, 
when $\phi_1$ and $\phi_2$ (\figref{fig:two_coins_mpg})
are selected to be updated
in the last iteration (multiple $\phi$'s can be updated in the same iteration when parallelizing VMP), 
$x_1$ and $x_2$ 
cannot be updated in the current iteration unfortunately 
because they require messages from $z_1$ and $z_2$, which were not selected and updated in the last iteration.
Working around this 
%
%if we have updated only 
%iteration, we can not update any of the other random variables because all of
%them have some neighbours that wasn't updated in the last iteration ($x$ and
%$\pi$ for $z$, $z$ for $\pi$ and $x$).  
%The only choice is to restart the
%Pregel operator whenever this happens.
through the use of primitive  \texttt{aggregateMessages} and \texttt{outerJoinVertices} API
 would not make life easier.
 Specifically, the user would have to 
 handle some low level details such as determining which intermediate RDDs 
 to insert to or evict from the cache.
 
% code
%a lot of branches in order to filter out vertices that should not receive the
%messages.

%
%
%
%
%second, 
%natural choice is Pregel API of GraphX.
%a vertex send messages to all its neighbor.
%describe above, there is a sequence.
%so,needs to add to do post-filtering.
%inefficient
%%
%to get around, fall back to
%can use aggregateMessages (like map, that parse the whole graphs/table as many aggregateMessage tasks).
%a lot of code to do pattern matching.
%and outjoinvertices (like reduce, that for the same key/vertex, the list of messages)

%Second, unlike some iterative machine learning algorithms, e.g., stochastic
%gradient descent (SGD), 
%that cache \emph{the base data} as RDD for repeated access,
%VMP inference would \emph{update} the values (e.g., the parameter) on the graph \emph{iteratively}
%and each iteration may access the \emph{updated graph of previous iterations} repeatedly.  
%When the user cannot cache all intermediate RDDs,
%she has to manually determine which intermediate RDDs should be insert to and evict from the cache.

Second, the user has to determine the best timing to do checkpointing so as to
avoid performance degradation brought by the long lineage created by many iterations.

Last but not the least, the user may have to customize a partition strategy for
each model being evaluated. 
GraphX built-in partitioning strategies are general and thus do not work well with message passing graphs,
which usually 
possess (i) complete bipartite components between the posteriors and the  observed variables
(e.g., $\phi_1$, $\phi_2$ and $x_1, \ldots, x_N$ in \figref{fig:two_coins_mpg}), and
(ii) large repetition of edge pairs induced from the plate (e.g.,  $N$ pairs of $\langle z_i, x_i\rangle$ in \figref{fig:two_coins_mpg}).
GraphX adopts a vertex-cut approach for graph partitioning
and a vertex would be replicated to multiple partitions if it lies on the cut.
So, imagine if the partition cuts on $x$'s in \figref{fig:two_coins_mpg}, 
that would incur large replication overhead as well as shuffling overhead.
Consequently, that really requires the domain users 
to have excellent knowledge on GraphX in order to carry out efficient inference on Spark. 
\section{InferSpark Overview}
\label{sec:framework}

\begin{figure*}[th]
	\centering
	\includegraphics[width=1.6\columnwidth]{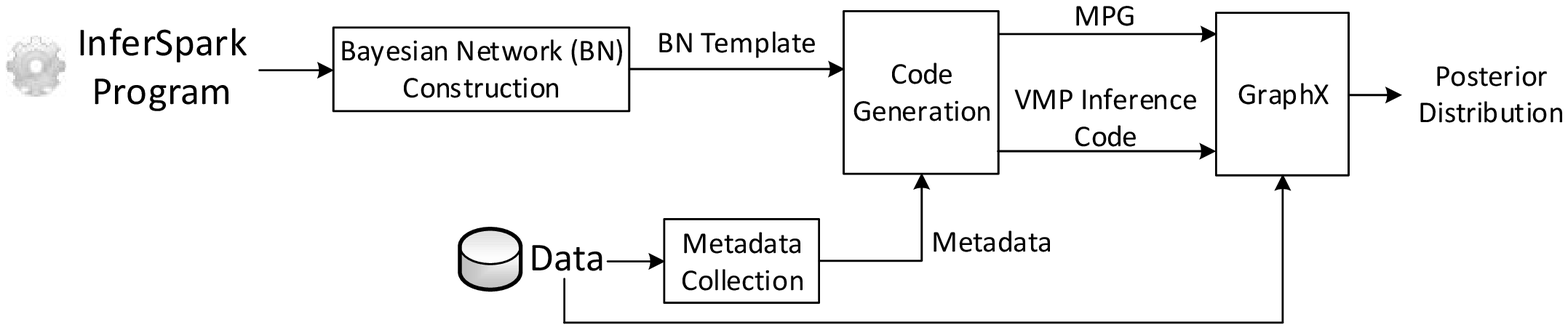}
	\caption{InferSpark Architecture}
	\label{fig:workflow}
\end{figure*}

%\KZ{My general feeling is that the running example is not made full use of.
%The discussion should be tightly coupled to the running example. E.g., when
%we talk about schedule, just present the schedule for the two coins.
%Some of the stuff here should go into implementation section.}

The overall architecture of InferSpark is shown in \figref{fig:workflow}. 
An InferSpark program is a mix of Bayesian network model definition and
normal user code. The Bayesian network construction module separates the
model part out, and transforms it into a Bayesian network template. This
template is then instantiated with parameters and meta data from the input
data at runtime by the code generation module, which produces the VMP inference
code and message passing graph. These are then executed on the GraphX 
distributed engine to produce the final posterior distribution.
%InferSpark analyzes the Bayesian network defined by a special 
%scala-like program and
%automatically transforms the model definition into the GraphX implementation
%of VMP algorithm. After two stages of compilation, the runtime system 
%launches the VMP implementation and returns the inference results 
%through the query API.  
Next, we describe the three key modules in more details with 
the example of the two-coin model (\figref{fig:two_coin_bn}).

\subsection{Running Example}

\begin{figure}[h]
\begin{lstlisting}
@Model class TwoCoins(alpha: Double, beta: Double) {
	val pi = Beta(alpha)
	val phi = (0L until 2L).map(_ => Beta(beta))
	val z = ?.map(_ => Categorical(pi))
	val x = z.map(z => Categorical(phi(z)))
}
object Main {
	def main() {
		val xdata: RDD[Long] = /* load (observed) data */
		val m = new TwoCoins(1.0, 1.0)
		m.x.observe(xdata)
		m.infer(steps=20)
		val postPhi: VertexRDD[BetaResult] = m.phi.getResult()
		/* postprocess */
		...
	}
}
\end{lstlisting}
\caption{Definition of two-coin model in InferSpark}
\label{fig:two_coins_modeldef}
\end{figure}

%Apart from ordinary scala code, the input program of InferSpark contains the
%statistical model definitions. The syntax of the model definition extends
%from the scala syntax.  
\figref{fig:two_coins_modeldef} shows the definition of the two-coin model
in InferSpark. The definition starts with ``{\sf @Model}'' annotation. 
The rest is similar to a class definition in
scala. The model parameters (``{\sf alpha}'' and ``{\sf beta}'') are constants to the
model. In the model body, only a sequence of value definitions are allowed,
each defining a random variable instead of a normal deterministic variable. 
The use of ``{\sf val}'' instead of ``{\sf var}'' in the syntax 
implies the conditional dependencies between random variables are fixed 
once defined. For example, line
2 defines the random variable $\pi$ having a symmetric Beta prior
$\mathrm{Beta}(\alpha, \alpha)$.

InferSpark model uses ``Range'' class in Scala to represent plates. Line 3
defines a plate of size 2 with the probabilities of seeing head in the 
two coins. The ``?'' is a special type of ``Range'' representing 
a plate of unknown size at the time of model definition. 
In this case, the exact size of the plate will be provided or inferred
from observed variables at run time.  When a random variable is
defined by mapping from a plate of other random variables, 
the new random variable is in the same plate as the others.  
For example, line 5 defines the outcomes $x$ as the mapping from $z$ 
to Categorical mixtures, therefore $x$ will be in the same plate as
$z$. Since the size of the plate surrounding $x$ and $z$ is unknown, we need
to specify the size at run time.  We can either explicitly set the length of
the ``?'' or let InferSpark set that based on the number of observed outcomes
$x$ (line 11).

\begin{figure}[th]
	\centering
	\includegraphics[width=0.25\textwidth]{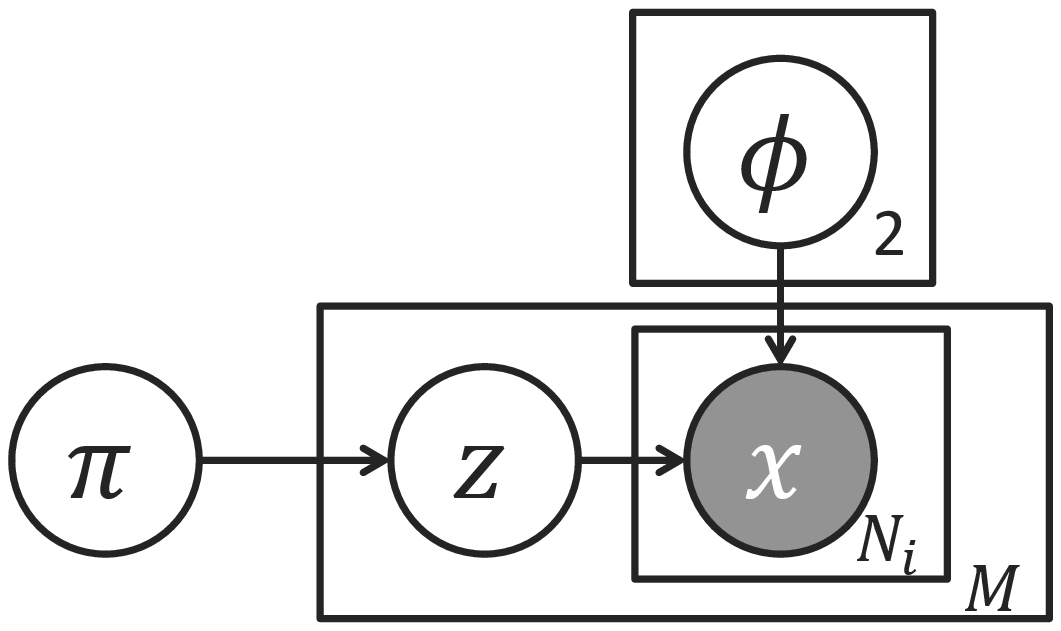}
	\caption{Two-coin Model with Nested Plates}
	\label{fig:two_coins_nestedplates}
\end{figure}

At the first glance, ``?'' seems redundant since it can be replaced by a
model parameter $N$ denoting the size of the plate.  However, ``?'' becomes
more useful when there are nested plates. In the two-coin model, suppose
after we choose one coin, we toss it multiple times. 
\figref{fig:two_coins_nestedplates} shows this scenario.
Then the outcomes $x$ are in two nested plates where the inner plate is
repeated $M$ times, and each instance may have
a different size $N_i$. Using the ``?'' syntax
for the inner plate, we simply change line 5 to
{\small\begin{verbatim}
	val x = z.map(z => ?.map(_ => Categorical(phi(z))))	
\end{verbatim}
}

\subsection{Bayesian Network Construction}

\begin{figure}[h]
\centering
\includegraphics[scale=0.4]{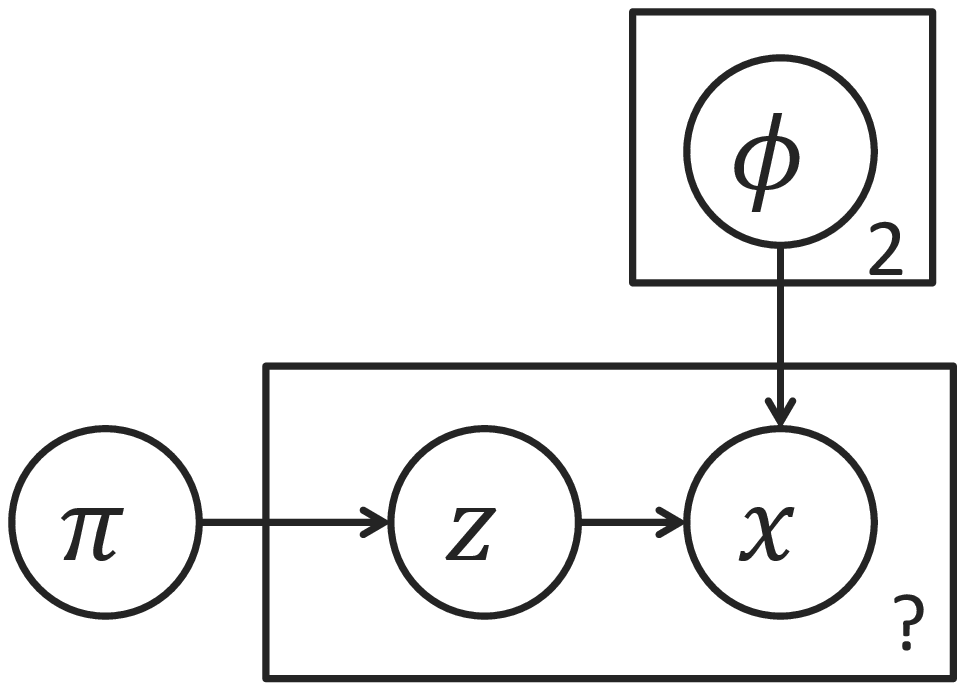}
\caption{Bayesian Network Template Constructed from the Two-coin Model}
\label{fig:two_coins_bn1}
\end{figure}

An input InferSpark program is first parsed and separated into two parts: 
the model definition (``{\sf @Model} class TwoCoins'' in 
\figref{fig:two_coins_modeldef}) and
the ordinary scala program (``{\sf object Main}'' in
\figref{fig:two_coins_modeldef}). The model definition is analyzed and
transformed into valid scala classes that define a Bayesian
network constructed from the model definition 
(e.g., \figref{fig:two_coins_bn1}) and the inference/query API.
Note the Bayesian network
constructed at this stage is only a template (different than 
\figref{fig:two_coin_bn}) because some of the information is not available 
until run time (e.g., the outcomes $x$, the number of coin flippings 
and the model parameters $\alpha$ and $\beta$). 

\subsection{Metadata Collection}

Metadata such as the observed values and the plate sizes missing from the
Bayesian networks are collected at runtime. In the two-coin
model, an instance of the model is created via the constructor invocation (e.g.
``{\sf val m = new TwoCoin(1.0, 1.0)}'' on line 10 of \figref{fig:two_coins_modeldef}). The constructor call provides
the missing constants in the prior distributions of $\pi$ and $\phi$. 
For each random variable defined in the model definition, 
there is an interface field with the
same name in the constructed object. Observed values are provided to InferSpark
by calling the ``{\sf observe}'' (line 11 of \figref{fig:two_coins_modeldef}) 
API on the field. 
There, the user provides an RDD of observed outcomes ``{\sf xdata}'' to InferSpark by calling
``{\sf m.x.observe(xdata)}''. The  {\sf observe} API also triggers 
the calculation of unknown plate sizes. 
In this case, the size of plate surrounding $z$ and $x$ is
automatically calculated by counting the number of elements in the RDD.

%When the user provide the observed random variables such as the words in the
%LDA model, the number of documents and the number of words in each document
%can be inferred from the data. This is different from most libraries in that
%they require the user to explicitly set the numbers or to transform the data
%into a library-specific format. InferSpark also tries to verify that the user
%have provided consistent data. For example, InferSpark will report an error,
%if the user provides data to both the topics $z$ and the words $w$ but they
%have differnt sizes.

\subsection{Code Generation}

\begin{figure}
\centering
	\includegraphics[width=0.35\textwidth]{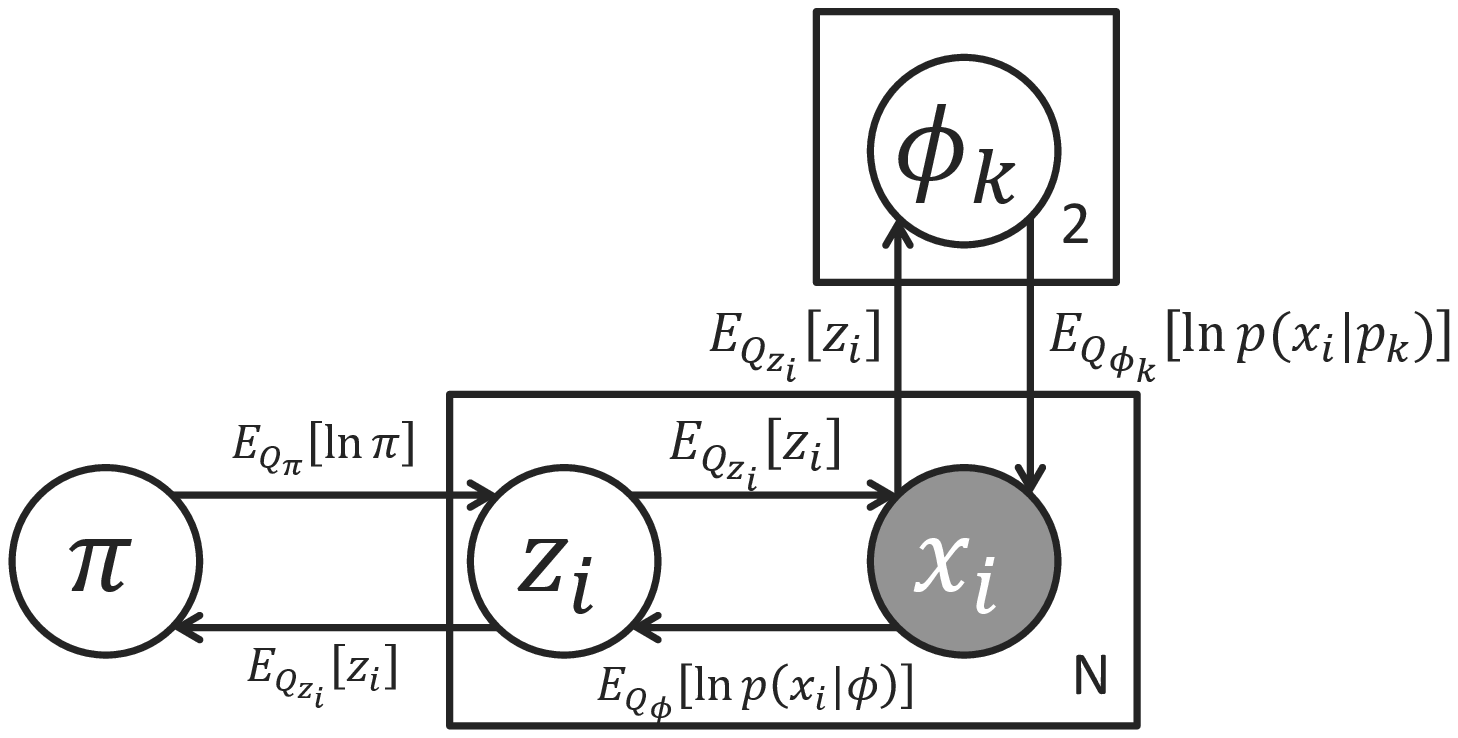}
	\caption{Bayesian Network with Messages for Two-coin Model}
	\label{fig:two_coins_msg}
\end{figure}

When the user calls ``{\sf infer}'' API (line 12 of \figref{fig:two_coins_modeldef}) 
on the model instance, InferSpark
checks whether all the missing metadata are collected. If so, it proceeds to
annotate the Bayesian network with messages used in VMP, resulting in
\figref{fig:two_coins_msg}. The expressions that calculate the messages 
(e.g., $E_{Q_\pi}[\ln \pi]$) depend on not only
the structure of the Bayesian network and whether the vertices are observed or
not, but also practical consideration of efficiency and constraints on
GraphX.  

To convert the Bayesian network to a message passing graph on GraphX,
InferSpark needs to construct a VertexRDD and an EdgeRDD. This step generates
the MPG construction code specific to the data.
\figref{fig:two_coins_mpg_constr_code} shows the MPG construction code
generated for the two-coin model. 
The vertices are constructed by the union
of three RDD's, one of which from the data and the others from 
parallelized collections (lines 8 and line 9 in \figref{fig:two_coins_mpg_constr_code}).
The edges are built from the data only. 
A partition strategy specific to the
data is also generated in this step.

\begin{figure}[h]
\begin{lstlisting}
class TwoCoinsPS extends PartitionStrategy {
	override def getPartition /**/
}
def constrMPG() = {
	val v1 = Categorical$13$observedValue.mapPartitions{
		initialize z, x */
	}
	val v2 = sc.parallelize(0 until 2).map{ /* initialize phi */ }
	val v3 = sc.parallelize(0 until 1).map{ /* initialize pi */ }
	val e1 = Categorical$13$observedValue.mapParititons{
		/* initialize edges */
	}
	Graph(v1 ++ v2 ++ v3, e1).partitionBy(new TwoCoinsPS())
}
\end{lstlisting}
\caption{Generated MPG Construction Code}
\label{fig:two_coins_mpg_constr_code}
\end{figure}

In addition to generating code to build the large message passing graph,
the codegen module also generates code for VMP iterative inference. 
InferSpark, which
distributes the computation, needs to create a schedule of parallel updates
that is equivalent to the original VMP algorithm, which only updates one vertex
in each iteration.  Different instances of the same random variables can be
updated at the same time. An example update schedule for the two-coins model is
($\pi$ and $\phi$) $\rightarrow$ $x$ $\rightarrow$ $z$ $\rightarrow$ $x$. VMP inference code that enforces the update
schedule is then generated. % \ERIC{how to derive this indeed?}

\subsection{Getting the Results}
%InferSpark finally compiles the VMP algorithm into a separate GraphX
%program and submit it to the Spark master. The user can specify how many
%iterations to run. 
The inference results can be queried through the ``{\sf getResult}''
API on fields in the model instance that retrieves a VertexRDD of approximate
marginal posterior distribution of the corresponding random variable. For
example, in Line 13 of \figref{fig:two_coins_modeldef}, ``{\sf m.phi.getResult()}'' 
returns a VertexRDD of two Dirichlet distributions. 
The user can also call ``{\sf lowerBound}'' 
on the model instance to get the evidence lower bound (ELBO) of the result, 
which is higher when the KL divergence between the approximate posterior 
distribution and the true posterior is smaller. 

\begin{figure}[h]
\centering
\begin{lstlisting}
var lastL: Double = 0
m.infer(20, { m =>
	if ((m.roundNo > 1) || 
		(Math.abs(m.lowerBound - lastL) < 
		   Math.abs(0.001 * lastL))) {
		false
	} else {
		lastL = m.lowerBound
		true	
	}
})
\end{lstlisting}
\caption{Using Callback function in ``infer'' API}
\label{fig:two_coins_callback}
\end{figure}

The user can also provide a callback function that will be called after
initialization and each iteration. In the function, the user can write
progress reporting code based on the inference result so far. 
For example, this function may return {\em false} whenever
the ELBO improvement is smaller than a threshold 
(see \figref{fig:two_coins_callback}) indicating the result is good enough 
and the inference should be terminated. 

%With all the information collected and calculated in previous steps,
%implementation of the inference algorithm as a GraphX program is generated,
%compiled and then executed. User can retrieve the posterior distributions as
%VertexRDDs through API calls.

%!TEX root = paper.tex
\section{Implementation}
\label{sec:implementation}
The main jobs of InferSpark are Bayesian network construction and  code generation (\figref{fig:workflow}).
Bayesian network construction first extracts Bayesian network template from the model definition and transforms it into a
Scala class with inference and query APIs at compile time. 
%The generated class
%is compiled with normal scala part of the program 
%Then, code generation takes those as inputs and generates a Spark program which 
%including the messaging passing graph and VMP inference code. Afterwards, the generated program is executed on Spark.
Then, code generation takes those as inputs and generates a Spark program
that can generate the messaging passing graph with VMP on top.
Afterwards, the generated program would be executed on Spark.

We use the code generation approach because it enables a more flexible API
than a library. For a library, there are fixed number of APIs for user to
provide data, while InferSpark can dynamically generate custom-made APIs 
according to the structure of the Bayesian network. 
Another reason for using code generation is
that compiled programs are always more efficient than interpreted programs.

%The code generation to implement InferSpark. 
%Stage 1 constructs
%a and submitted to Spark.  
%Stage 2 performs metadata collection using Spark and generate MPG 
%construction and VMP inference code based on the meta data at run time.

%\subsection{Stage 1 Code Generation}
\subsection{Bayesian Network Construction}\label{bnc}

In this offline compilation stage, the model definition is first transformed into a Bayesian network.
We use the macro annotation, a compile-time meta programming facility of
Scala.  It is currently supported via the macroparadise plugin. After the
parser phase, the class annotated with ``{\sf @Model}'' annotation is passed from the
compiler to its transform method. InferSpark treats the class passed to it as
model definition and transforms it into a Bayesian network.

\begin{figure}[!h]
\scriptsize
	\begin{tabular}{lrl}
		ModelDef		& ::= & `@Model' `class' id \\
					&     &`(' ClassParamsOpt `)' `\{' Stmts `\}' \\
		ClassParamsOpt	& ::= & `' /* Empty */ \\
						&	| &	ClassParams \\
		ClassParams		& ::= & ClassParam  [`,' ClassParams] \\
		ClassParam		& ::= & id `:' Type \\
		Type			& ::= & `Long' | `Double' \\
		Stmts			& ::= & Stmt [[semi] Stmts]\\
		Stmt			& ::= & `val' id = Expr \\
		Expr			& ::= & `\{' [Stmts [semi]] Expr `\}' \\
						&	| & DExpr	\\
						&   | & RVExpr \\
						&	| & PlateExpr \\
						&	| & Expr `.' `map' `(' id => Expr `)'\\
		DExpr			& ::= & Literal	\\
						&   | & id \\
						&   | & DExpr (`+' | `-' | `*' | `/') DExpr \\
						&   | & (`+' | `-') DExpr	\\
		RVExpr			& ::= & `Dirichlet' `(' DExpr `,' DExpr `)' \\
						&   | & `Beta' `(' DExpr `)' \\
						&   | & `Categorical' `(' Expr `)' \\
						&   | & RVExpr RVArgList	\\
						&   | & id	\\
		RVArgList		& ::= & `(' RVExpr `)' [ RVArgList ] \\
		PlateExpr		& ::= & DExpr `until' DExpr	\\ 
						&   | & DExpr `to' DExpr	\\
						&	| & `?' \\
						&	| & id
	\end{tabular}
\caption{InferSpark Model Definition Syntax}
\label{fig:inferspark_syntax}
\end{figure}

\figref{fig:inferspark_syntax} shows the syntax of InferSpark model
definition. The expressions in a model definition is divided into 3
categories: deterministic expressions (DExpr), random variable expressions
(RVExpr) and plate expressions. The deterministic expressions include
literals, class parameters and their arithemetic operations. The random
variable expressions define random variables or plates of random variables.
The plate expressions define plate of known size or unknown size. The random
variables defined by an expression can be binded to an identifier by the value
definition. It is also possible for a random variable to be binded to multiple
or no identifiers. To uniquely represent the random variables, we assign
internal names to them instead of using the identifiers.

\begin{figure}[ht]
\centering
	\includegraphics[width=0.6\columnwidth]{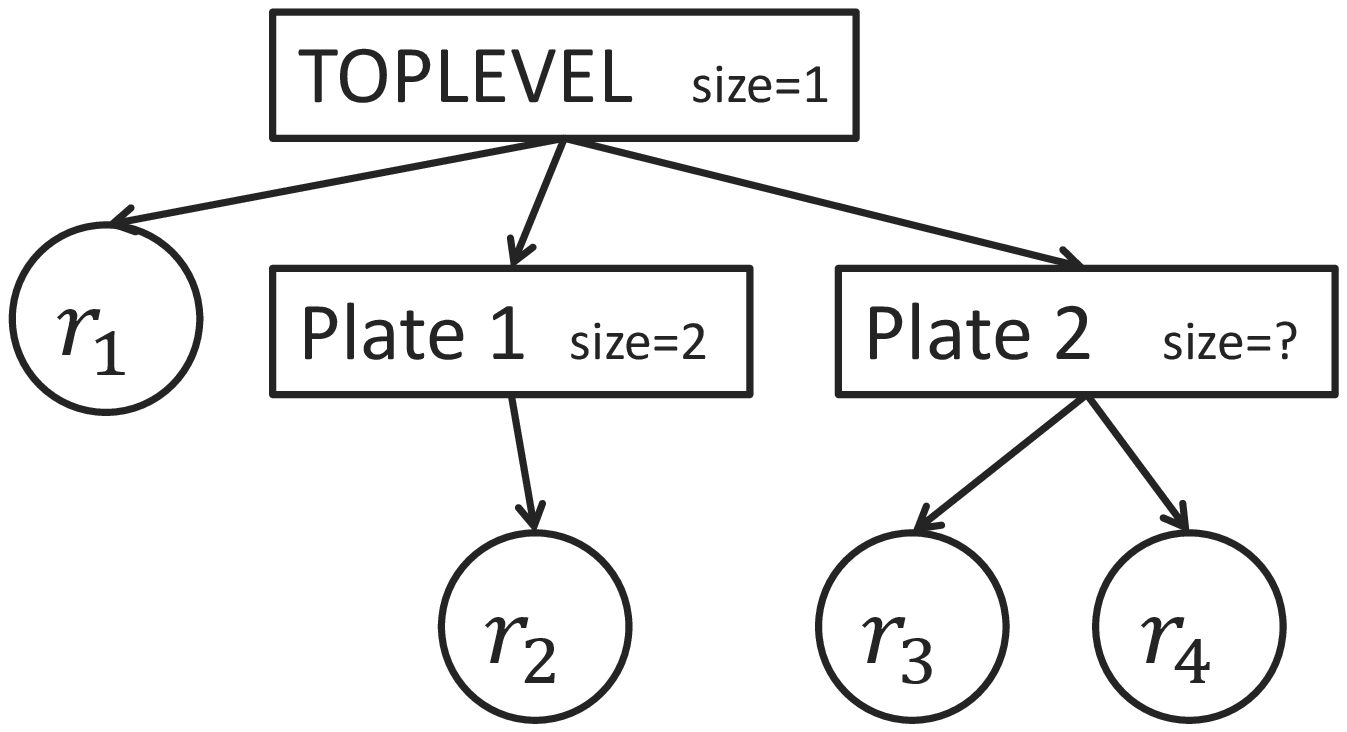}
	\caption{Internal Rep. of Bayesian Network}
	\label{fig:two_coins_internal_bn1}
\end{figure}

Internally, InferSpark represents a Bayesian network in a tree form, where the
leaf nodes are random variables and the non-leaf nodes are plates. The edges
in the tree represent the nesting relation between plates or between a plate
and random variables. The conditional dependencies in the Bayesian network are
stored in each node.  The root of the tree is a predefined plate TOPLEVEL with
size 1.  \figref{fig:two_coins_internal_bn1} is the internal representation of
the two-coin model in \figref{fig:two_coins_bn1}, where
$r_1$, $r_2$, $r_3$, $r_4$, correspond to $\pi$, $\phi$, $z$, $x$, respectively. 
Plate 1 and Plate 2 correponds to the plates defined on lines 3--5 in
\figref{fig:two_coins_modeldef}. 

If a plate is nested within another plate,
the inner plate is repeated multiple times, in which case,
the size attribute of the plate node will be computed by summing the size of each
repeated inner plate. We call the size attribute in the tree {\em flattened size}
of a plate. For example, in \figref{fig:two_coins_nestedplates}, the flattened
size of the innermost plate around $x$ is $\sum_i N_i$.

InferSpark recursively descends on the abstract syntax tree (AST) of the model definition to construct
the Bayesian network.   In the
model definition, InferSpark follows the normal lexical scoping rules.
%\figref{fig:transformation_rules} shows part of the transformation rules of each
%syntax. ``transform(ast, curPlate, env)'' transforms non-expression parts while
%``eval(ast, curPlate, env)'' transforms expression and result the evaluation result.
InferSpark evaluates the expressions to one of the following three results
\begin{itemize}
	\item a node in the tree
	\item a pair $(r, \textrm{plate})$ where $r$ is a random variable node
		and plate is a plate node among its ancestors, which represents all
		the random variables in the plate
	\item a determinstic expression that will be evaluated at run time
\end{itemize}
%When defining the random variables, InferSpark also checkes the conjugacy
%constraints of the Bayesian network.

At this point, apart from constructing the Bayesian network representation, 
InferSpark also generates the code for metadata collection, a module used in 
stage 2. For each random variable name bindings, a singleton interface object 
is also created in the resulting class. 
The interface object provides ``{\sf observe}'' and ``{\sf getResult}'' API for later use.

\subsection{Code Generation}

Code generation happens at run time. It is divided into 4 steps: metadata
collection, message annotation, MPG construction code generation and inference
execution code generation.

Metadata collection aims to collect the values of the model parameters,
check whether random variables are observed or not, the flattened sizes of the plates.
These metadata can help to 
assign VertexID to the vertices on the message passing graph.  
%If a plate is the innermost plate that a random variable is in, the flat size of
%the plate is defined as the number of this random variable. For example, the
%flat size of plate 2 in \figref{fig:two_coins_internal_bn1} is the number of
%outcomes $x$ or the number of the random variable $z$.  
After the flattened sizes of plates are calculated, we can assign VertexIDs to the
vertices that will be constructed in the message passing graph. Each random
variable will be instantiated into a number of vertices on the MPG where the
number equals to the flattened size of its innermost plate. The vertices
of the same random variable are assigned consecutive IDs. For example, $x$ may
be assigned ID from $0$ to $N-1$. The intervals of IDs of random variables in
the same plate are also consecutive. A possible ID assignment to $z$ is $N$ to
$2N - 1$. Using this ID assignment, we can easily i) determine which random
variable the vertex is from only by determining which interval the ID lies
in; ii) find the ID of the corresponding random variable in the same plate by
substracting or adding multiples of the flattened plate size (e.g. if $x_i$' ID is
$a$ then $z_i$'s ID is $a + N$).

Message annotation aims to annotate the Bayesian Network Template from the previous stage (Section \ref{bnc})
with messages
to be used in VMP algorithm.  The annotated messages are stored in the form of
AST and will be incorporated into the the generated code, output of this stage. 
The rules of the messages to annotate are predefined according to the
derivation of the VMP algorithm.
%The messages between different types of random variables without mixture can
%be created by looking up in a table. For a mixture variable, the message from
%it is composed from basic messages in the table. 
%\KZ{Where is the table, what
%is it like?}
%
%We also need to generate two additional functions for each random variable: a
%message merge function and a vertex update function. 
%\KZ{Don't understand:
%The messages are first
%sent to the vertices to update and merged on the fly, then joined with the
%original vertices to apply the update.}  
After the messages are generated, we
generate for each type of random variable a class with the routines for
calculating the messages and updating the vertex. 

The generated code for constructing  the message passing graph requires  building a VertexRDD
and an EdgeEDD. The VertexRDD is an RDD of VertexID and vertex attribute pairs.
Vertices of different random variables
are from different RDDs (e.g., {\sf v1}, {\sf v2}, and {\sf v3} in \figref{fig:two_coins_mpg_constr_code})
and have different initialization methods.
For unobserved random variables, the source can be any RDD that has the same
number of elements as the vertices instantiated from the random variable. For
observed random variables, the source must be the data provided by the user. If
the observed random variable is in an unnested plate, the vertex id can be
calculated by first combining the indices to the data RDD then adding an offset.

One optimization of constructing the EdgeRDD is to \emph{reverse the edges}.
If the code generation process generates an EdgeRDD in straightforward manner,
the {\sf aggregateMessages} function 
has to scan all the edges 
to find edges whose destinations are of $v$ type
because GraphX indexes the \emph{source} but not the \emph{destination}.
Therefore, when constructing the EdgeRDD, we generate code that reverses the edge
so as to enjoy the indexing feature of GraphX.
When constructing the graphs,
we also take into account the graph partitioning scheme 
because that has a strong influence on the performance.
We discuss this issue in the next section.

The final part is to generate the inference execution code that implements the
iterative update of the VMP algorithm. 
We aim to generate code that updates each vertex in the
message passing graph at least once in each iteration. 
As it is safe to update vertices that do not
have mutual dependencies, i.e., those who do not send messages to one another,
we divide each iteration into substeps.
Each substep updates a portion of the
message passing graph that does not have mutual dependencies. 

A substep in each iteration consists of two GraphX operations:
{\sf aggregateMessages} and {\sf outerJoinVertices}. Suppose {\sf g} is the message passing
graph, the code of a substep is:
\begin{lstlisting}
val prevg = g
val msg = g.aggregateMessages(sendMsg, mergeMsg, TripletFields)
g = g.outerJoinVertices(msg)(updateVertex).persist()
g.edges.count()
prevg.unpersist()
\end{lstlisting}

The RDD {\sf msg} does not need to be cached because it is only used once. But
the code generated has to cache the graph {\sf g} because the graph is used 
twice in both {\sf aggregateMessages} and {\sf outerJoinVertices}. However, only caching it is not
enough, the code generation has to include a line like 4 above to activate the caching process.
Once {\sf g} is cached, code generation evicts the previous (obsolete) graph {\sf prevg} from the cache.
To avoid the long lineage caused by iteratively updating message passing graph, which will overflow the heap space of the drive,
the code generation process also adds a line of code to checkpoint the graph to HDFS 
every $k$ iterations.

\subsection{Execution}

The generated code at run time are sent to the Scala compiler. The resulting
byte code are added to the classpath of both the driver and the workers. Then
InferSpark initiates the inference iterations via reflection invocation.

\subsection{Discussion on Partitioning Strategies}

\begin{figure}[h]
	\includegraphics[width=0.45\textwidth]{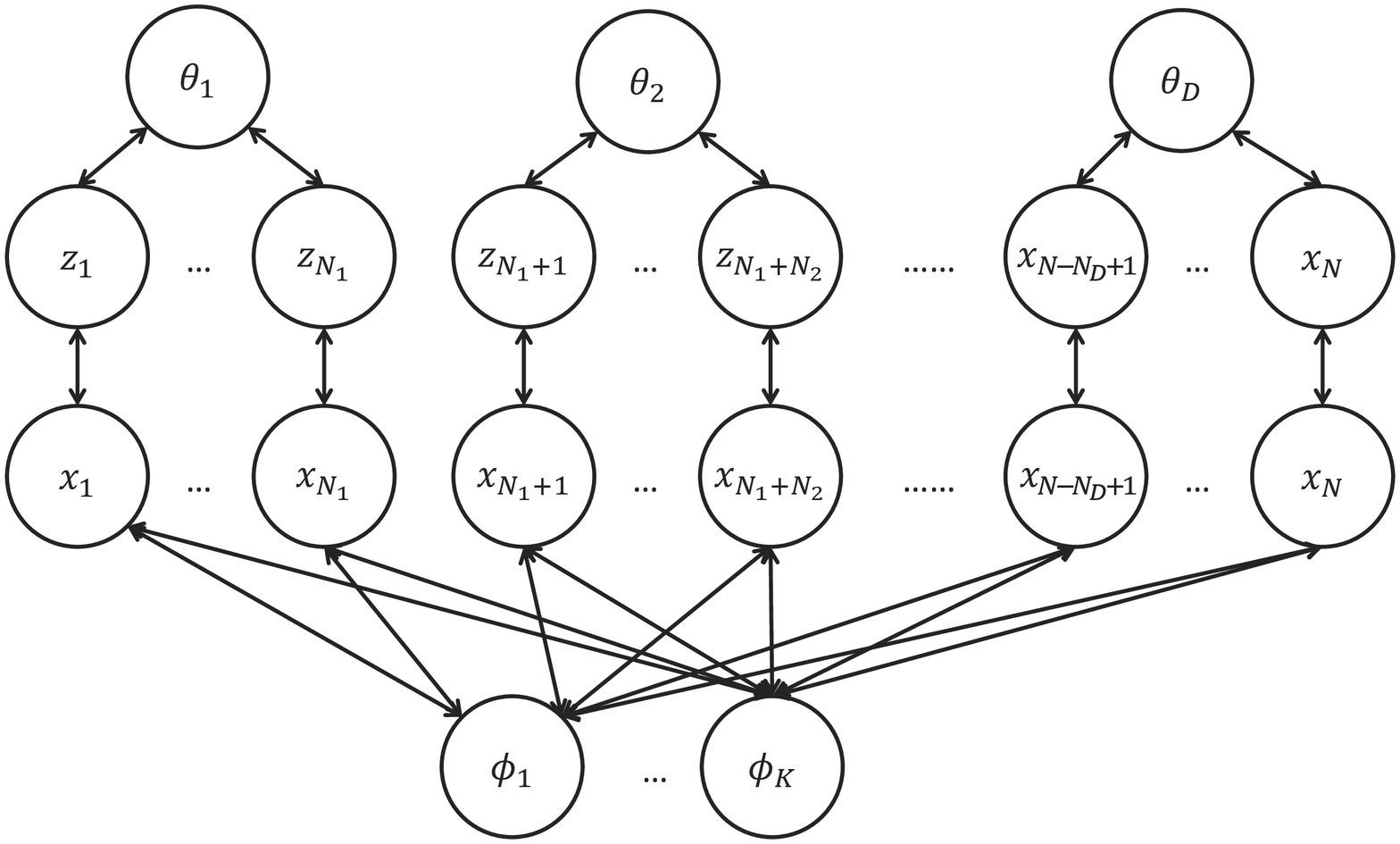}
	\caption{Message Passing Graph of a Mixture Model}
	\label{fig:mixture_mpg}
\end{figure}

GraphX adopts a vertex-cut partitioning approach.
The vertices are replicated in edge partitions instead of edges being replicated in vertex
partitions. 
%This replication creates a triplet view, which is a 3-tuple of
%source vertex, edge and destination vertex. The triplet view is not
%materialized until referenced. For our purpose, only {\sf aggregateMessages}
%use the destination end of the triplet view, so only the destination of each
%edge is replicated in the edge partitions. We want to minimize the number of
%replications so that outerJoin is more efficient. 
The four built-in partition
strategies in GraphX are: 
EdgePartition1D (1D),
EdgePartition2D  (2D),
RandomVertexCut (RVC), and
CanonicalRandomVertexCut (CRVC).
In the following, we first show that these general 
partitioning strategies perform badly for the VMP algorithm on MPG.
Then, we introduce our own partitioning strategy.

\figref{fig:mixture_mpg} shows a more typical message passing graph of a
mixture model instead of the toy two-coin model that we have used so far. $N$
is the number of $x$ and $z$, $K$ is the number of $\phi$, $D$ is the number
of $\theta$. Typically, $N$ is very large because that is the data size (e.g.,
number of words in LDA), $K$ is a small constant (e.g., number of topics in
LDA), and $D$ could be a constant or as large as $N$ (e.g., number of
documents in LDA).

EdgePartition1D essentially is a random partitioning strategy, except that it
co-locates all the edges with the same source vertex. Suppose all the edges from
$\phi_k$ are assigned to partition $k$. Since there's an edge from $\phi_k$ to
each one of the $N$ vertices $x$, partition $k$ will have the replications
of all $x_1, x_2, \ldots, x_N$. In the best case, 
edges from different $\phi_k$ are assigned to different
partitions. Then the largest edge partition still have at least $N$ vertices.
When $N$ is very large, the largest edge partition is also very large, which
will easily cause the size of an edge partition to exceed the RDD block size limit. However,
the best case turns out to be the worst case 
when it comes to the number of vertex replications
because it actually replicates the size $N$ data $K$ times, which is
extremely space inefficient. The over-replication also incurs large amount of
shuffling when we perform outer joins because each updated vertex has to
be shipped to every edge partition, prolonging the running time. 

We give a more formal analysis of the number of vertices in the largest edge
partition and the expected number of replications of $x_i$ under
EdgePartition1D. As discussed above, there's at least one edge partition that
has replications of all the $x_i$'s. 
Observe that the graph has an upper bound of
$3N + K$ vertices, so the number of vertices in the largest edge partition is
$O(N)$. Let $N_{x_i}$ be the number of replications of $x_i$, then the expected
number of replication of $x_i$ is 
\begin{align*}
	E[N_{x_i}] &= M(1 - (1 - \frac{1}{M})^{K+1}) \\
		&= \left\{
			\begin{array}{ll}
				(K + 1) + o(1) & K = O(1) \\
				M + o(1) & K = O(M) 
			\end{array}
		\right.%}
\end{align*}

%first introduction 
%then analysis
\begin{figure}[h]
	\centering
	\includegraphics[scale=0.3]{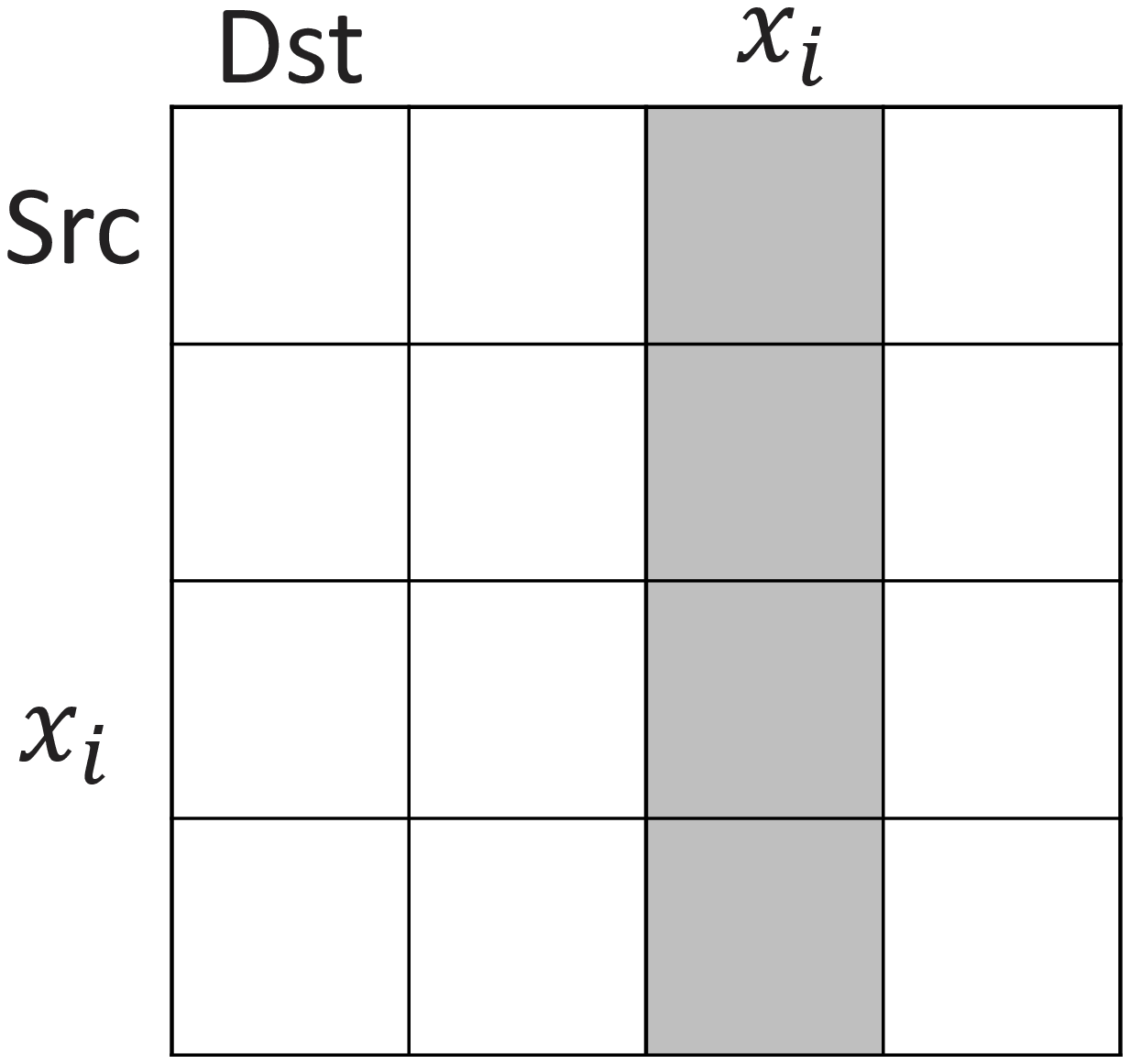}
	\caption{EdgePartition2D.  Each grid is a partition. The possible partitions in which $x_i$ is
	replicated is shaded}
	\label{fig:2dhash}
\end{figure}

EdgePartition2D evenly divides the adjacency matrix of the graph into $\sqrt{M}
\times \sqrt{M}$ partitions. The vertices are uniformly distributed along the
edges of the adjacency matrix by hashing into $\sqrt{M}$ buckets. The upper
bound of the number of replications of a vertex $x_i$ is $\sqrt{M}$ because all
the edges that point to it are distributed to at most $\sqrt{M}$ partitions
in shown as \figref{fig:2dhash}.  Meanwhile, there are $K+1$
edges pointing to $x_i$, so the number of replications of $x_i$ cannot exceed
$K+1$ as well. Therefore, the upper bound of replications of $x_i$ is actually
$\min(K+1, \sqrt{M})$. On the other hand, suppose each of the $\phi_k$ is
hashed to different bucket and $N$ $x$'s are evenly distributed into the
$\sqrt{M}$ buckets, then the number of largest partition is at least
$\frac{N}{\sqrt{M}}$, which is still huge when the average number of words per
partition is fixed. Following is the formal analysis of the EdgePartition2D.

Let $B$ be an arbitrary
partition in the dark column on \figref{fig:2dhash}.
Let $Y_{x_i, B}$ be the indicator variable for the event that $x_i$ is replicated in 
Then the expectation of $Y_{x_i, B}$ is
\begin{align*}
	E[Y_{x_i, B}] &= 1 - (1 - \frac{1}{\sqrt{M}})^{K+1} \\
\end{align*}

The number of vertices $N_B$ in the largest partition $B$ is at least the expectation
of the number of vertices in a partition, which is also at least the
expectation of the number of $x_i$ in it:
\begin{align*}
	E[N_{B}] &= \sum_{v} E[Y_{v, B}] \\
		&\ge \frac{N}{\sqrt{M}} E[Y_{x_i, B}]	\\
		& = \left\{
				\begin{array}{ll}
					(K + 1)\eta	+ o(1) & K = O(1) \\
					\sqrt{M}\eta + o(1) & K = O(M) 
				\end{array}
			\right.%}
\end{align*}

The expected number of replications of $x_i$ is
\begin{align*}
	E[N_{x_i}] &= \sqrt{M}E[Y_{x_i, B}] \\
		&= \left\{
			\begin{array}{ll}
				(K + 1) + o(1) & K = O(1) \\
				\sqrt{M} + o(1) & K = O(M) 
			\end{array}
		\right.%}
\end{align*}

RandomVertexCut (RVC) uniformly assigns each edge to one of the $M$
partitions. The expected number of replications of $x_i$ tends to be $O(K)$
when $K$ is a constant and tends to be $O(N)$ when $K$ is proportional to the
number of partitions. The number of vertices in the largest partition is also
excessively large. It is $O(K\frac{N}{M})$ when K is a constant and $O(N)$
when $K$ is proportional to the number of partitions.  CanonicalRandomVertexCut
assigns two edges between the same pair of vertices with opposite directions
to the same partition. For VMP, it is the same as RandomVertexCut since only
the destination end of an edge is replicated. For example, if $x_i$ has $K +1$
incoming edges, then the probability that $x_i$ will be replicated in a
particular partition is independent from whether edges in opposite
direction are in the same partition or randomly distributed. Therefore 
CRVC will have the same result as RVC.  
\tabref{tab:max_v_per_edge_part_O1} and
\tabref{tab:max_v_per_edge_part_OM} summarize the comparison of different
partition strategies.

InferSpark's partitioning strategy is actually tailor-made for VMP's message passing graph.
The intuition is that the MPG has a special structure.
For example, in \figref{fig:mixture_mpg},
we see that the MPG essentially has $D$ ``independent'' trees rooted at $\theta_i$,
where the leaf nodes  are $x$'s and they form a complete bipartite graph with all $\phi$'s.
In this case, one good partitioning strategy is to form $D$ partitions, 
with each tree going to one partition and the $\phi$'s getting replicated $D$ times.
We can see that such a partition strategy incurs no replication on $\theta$, $z$, and $x$,
and incurs $D$ replications on $\theta$.

Generally, our partitioning works as follows: Given an edge, we first determine
which two random variables (e.g. $x$ and $z$) are connected by the edge. It is
quite straightforward because we assign ID to the set of vertices of the same
random variable to a consecutive interval. We only need to look up which
interval it is in and what the interval corresponds to. Then we compare the
total number of vertices correponding to the two random variables and choose
the larger one. Let the Vertex ID range of the larger one to be $L$ to $H$. We
divide the range from $L$ to $H$ into $M$ subranges. The first subrange is $L$
to $L + \frac{H-L+1}{M}$; the second is $L + \frac{H-L+1}{M} + 1$ to $L +
2\frac{H-L+1}{M}$ and so on. If the vertex ID of the edge's chosen vertex falls
into the $m^{th}$ subrange, the edge is assigned to partition $m$.

In the mixture case, at least one end of every edge is $z$ or $x$. Since
the number of $z$'s and $x$'s are the same, 
each set of edges that link to the $z_i$ or $x_i$ with 
the same $i$ are co-located. This guarantees that $z_i$
and $x_i$ only appears in one partition. All the $\phi_k$'s are replicated in each
of the $M$ partitions as before. The only problem is that many $\theta_j$ with
small $N_j$ could be replicated to the same location. In the worst case, the
number of $\theta$ in one single partition is exactly $\eta$. However, it is
not an issue in that case because the number of vertices in the largest
partition is still a constant $3\eta + K$. It is also independent from whether $K
= O(1)$ or $K = O(M)$.

\begin{table}[h]
	\centering
	\caption{Analysis of Different Partition Strategies When $K = O(1)$}
	\label{tab:max_v_per_edge_part_O1}
	\begin{tabular}{lll}
		\hline
		Partition Strategy & $E[N_{x_i}]$ & $E[N_B]$\\\hline\hline
		1D & $O(K)$ & $O(N)$ \\\hline
		2D & $O(K)$ & $O(K\frac{N}{M})$ \\\hline
		RVC & $O(K)$ & $O(K\frac{N}{M})$ \\\hline
		CRVC & $O(K)$ & $O(K\frac{N}{M})$ \\\hline
		{\bf InferSpark} & 1 & $3\frac{N}{M}+1$ \\\hline
	\end{tabular}
\end{table}

\begin{table}[h]
	\centering
	\caption{Analysis of Different Partition Strategies When $K = O(M)$}
	\label{tab:max_v_per_edge_part_OM}
	\begin{tabular}{lll}
		\hline
		Partition Strategy & $E[N_{x_i}]$ & $E[N_B]$\\\hline\hline
		1D & $O(M)$ & $O(N)$ \\\hline
		2D & $O(\sqrt{M})$ & $O(\sqrt{M}\frac{N}{M})$ \\\hline
		RVC & $O(M)$ & $O(N)$ \\\hline
		CRVC & $O(M)$ & $O(N)$ \\\hline
		{\bf InferSpark} & 1 & $3\frac{N}{M}+1$ \\\hline
	\end{tabular}
\end{table}

\section{Evaluation}
\label{sec:eval}

\begin{figure*}
\centering
    \subfigure[LDA]{
        \label{fig:exp_lda}
        \includegraphics[width=0.3\linewidth]{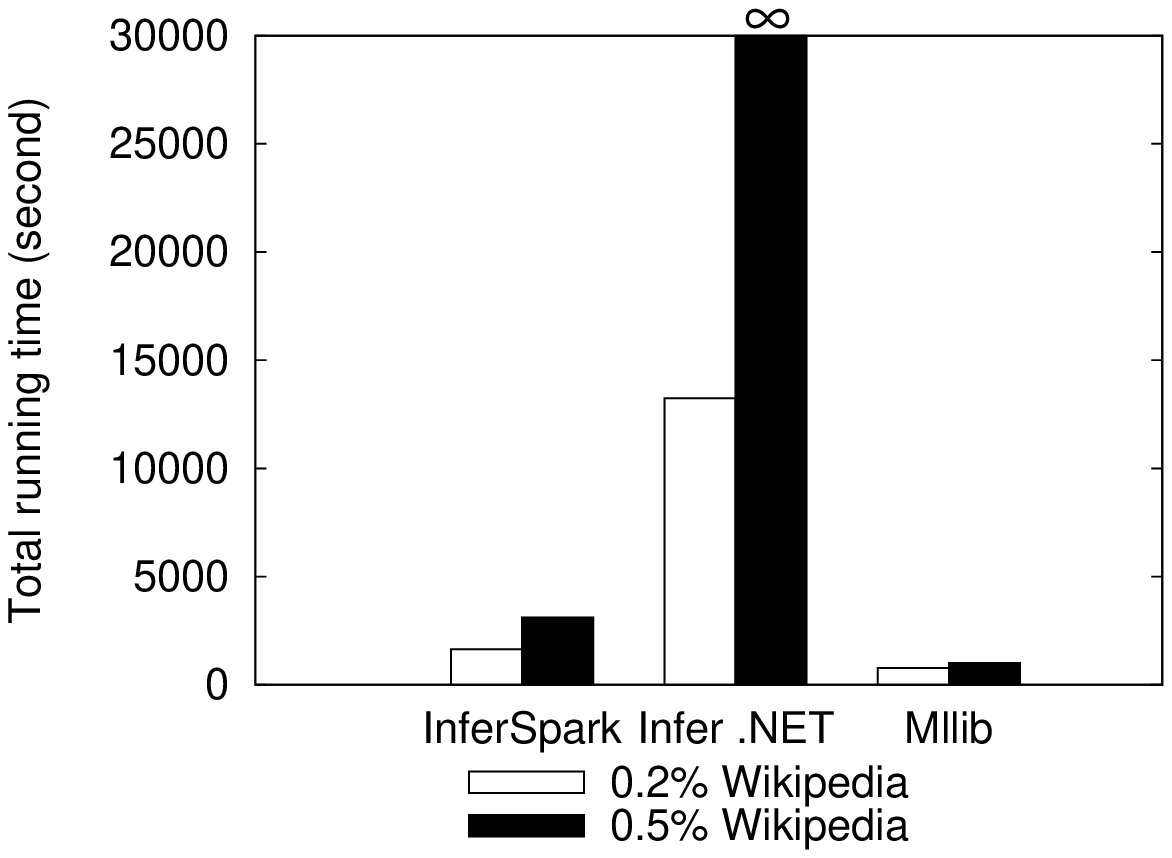}
    }
    \subfigure[SLDA]{
        \label{fig:exp_slda}
        \includegraphics[width=0.3\linewidth]{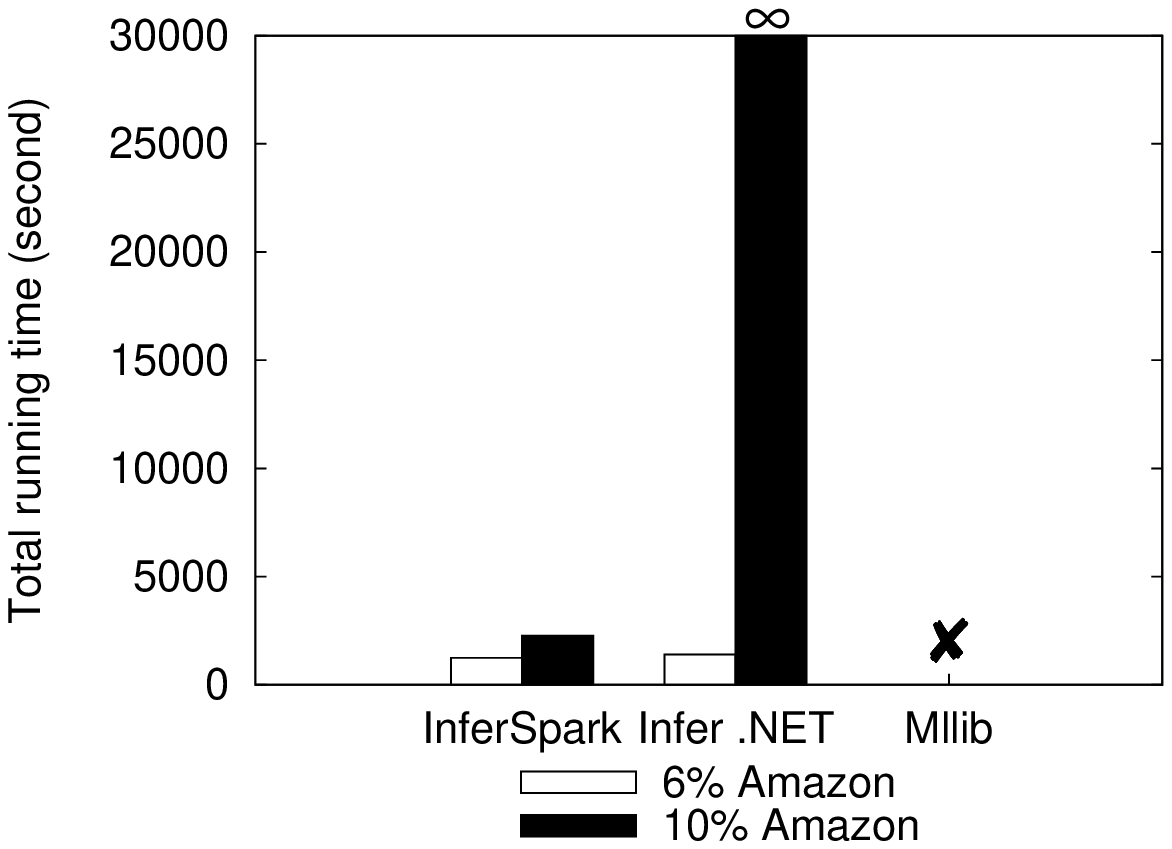}
    }
    \subfigure[DCMLDA]{
        \label{fig:exp_dcmlda}
        \includegraphics[width=0.3\linewidth]{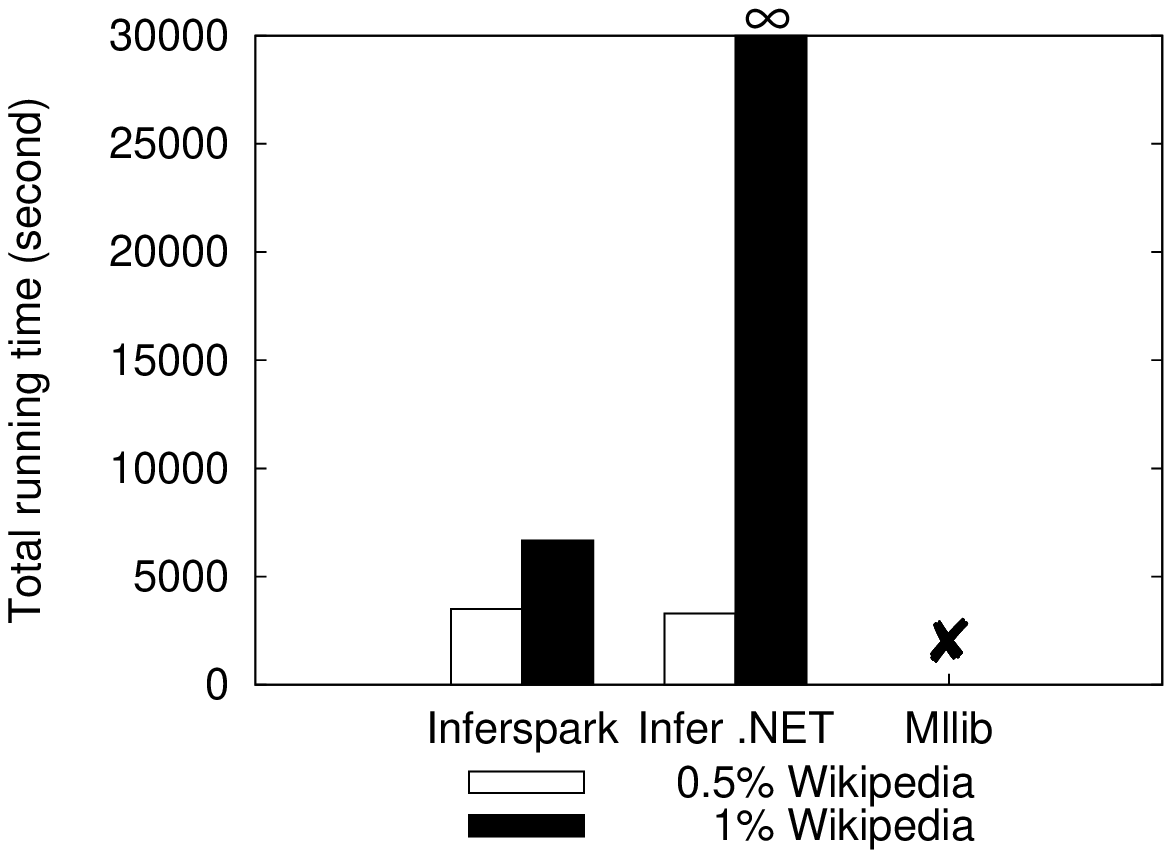}
    }
	\caption{Running Time}
    \label{fig:exp_comparison}
\end{figure*} 

In this section, we present performance evaluation of InferSpark, based on constructing 
and carrying out statistic inference on three models: Latent Dirichlet Allocation (LDA), Sentence-LDA (SLDA) \cite{Jo2011}, and Dirichlet Compound Multinomial LDA (DCMLDA) \cite{Doyle2009}.
LDA is a standard model in topic modeling, which takes in a collection of documents
and infers the topics of the documents.  
Sentence-LDA (SLDA) is a model for finding aspects in online reviews, which takes in online reviews, and infers the aspects.
Dirichlet Compound Multinomial LDA (DCMLDA) is another topic model that accounts for burstiness in documents.
All models can be implemented in InferSpark using less than 9 lines of code (see \figref{fig:intro_lda_def} and Appendix \ref{models}).
For comparison, we include MLlib in our study whenever applicable.
MLlib includes LDA as standard models.  However, MLlib does not include SLDA and DCMLDA.
There are other probabilistic programming frameworks apart from Infer.NET (see Section \ref{sec:related}).
All of them are unable to scale-out onto multiple machines yet.  
Infer.NET so far is the most predominant one with the best performance, so we also include it in our study whenever applicable.

All the experiments are done on nodes running Linux with 2.6GHz quad-core, 32GB memory, and 700GB hard disk. 
Spark 1.4.1 with scala 2.11.6 is installed on all nodes.
The default cluster size for InferSpark and MLlib is 24 data nodes and 1 master node.
Infer.NET can only use one such node.
The data for running LDA, SLDA, and DCMLDA are listed in Table \ref{data}.
The wikipedia dataset is the wikidump. Amazon is a dataset of Amazon reviews
used in \cite{Jo2011}.
We run 50 iterations and do checkpointing every 10 iterations for each model on each dataset.

\begin{table}\scriptsize
\caption{Datasets}
\label{data}
\begin{tabular}{ccc}
     \begin{tabular}{|c|c|c|}     \hline
        {Wikipedia} & words & topics \\\hline\hline
         0.2\% & 541,644 & 96 \\\hline
         0.5\% & 1,324,816  & 96 \\         \hline
         \multicolumn{3}{c}{LDA} \\
     \end{tabular}
          &
     \begin{tabular}{|c|c|c|}     \hline
        {Amazon} & words & topics \\\hline\hline
         6\% & 349,569 & 96 \\\hline
         10\% & 607,430  & 96 \\         \hline
         \multicolumn{3}{c}{SLDA} \\
     \end{tabular}
     \\\\
     \multicolumn{2}{c}{
     \begin{tabular}{|c|c|c|}     \hline
        {Wikipedia} & words & topics \\\hline\hline
         0.5\% & 1,324,816 & 10 \\\hline
         1\% & 2,596,155  & 10 \\         \hline
         \multicolumn{3}{c}{DCMLDA} \\
     \end{tabular}
     }
\end{tabular}
\end{table}

\subsection{Overall Performance}

\figref{fig:exp_comparison} shows the time of running LDA, SLDA, and DCMLDA 
on InferSpark, Infer.NET, and MLlib.
Infer.NET cannot finish the inference tasks on all three models within a week.
MLlib supports only LDA, and is more efficient than InferSpark in that case.
However, we remark that MLlib uses the EM algorithm which only
calculates Maximum A Posterior instead of the full posterior and is specific to LDA.
In contrast, InferSpark aims to provide a handy programming platform for statistician and domain users to build and test various customized models based on big data.
It would not be possible to be done by any current probabilistic frameworks nor with Spark/GraphX directly unless huge programming effort is devoted.  
MLlib versus InferSpark 
is similar to C++ programs versus DBMS: highly optimized C++ programs are more efficient, 
but DBMS achieves good performance with lower development time.
From now on, we focus on evaluating the performance of InferSpark.

Table \ref{breakdown} shows the time breakdown of InferSpark.
The inference process executed by GraphX, as expected, dominates the running time.
The MPG construction step executed by Spark, can finish within two minutes.
The Bayesian network construction and code generation can be done in seconds.

\begin{table*}
\caption{Time Breakdown (in seconds and \%)}
\label{breakdown}
\begin{tabular}{|l||*{8}{r|}r|}
\hline
Model & \multicolumn{2}{c|}{B.N. Construction} & \multicolumn{2}{c|}{Code Generation}	& \multicolumn{4}{c|}{Execution} & Total \\\cline{6-9} 
  & \multicolumn{2}{c|}{ } & \multicolumn{2}{c|}{ }	& \multicolumn{2}{c|}{MPG Construction} & \multicolumn{2}{c|}{Inference} &	 \\ \hline \hline
LDA 541644 words	& 21.911	& 1.34\%	& 11.15 &	0.68\%	& 38.147	& 2.33\% &	1566.692 & 95.65\%	& 1637.9 \\ \hline
LDA 1324816 words &	21.911 & 0.70\% & 12.25	& 0.39\% & 79.4 & 2.55\%	& 3002.1 & 96.36\% &	3115.661 \\ \hline
SLDA 349569 words &	21.867 & 1.76\% & 11.05 &	0.89\%	& 26.33 & 2.12\% &	1182.2	& 95.23\%	& 1241.447 \\ \hline
SLDA 607430 words & 21.867 & 0.96\%	& 11.69	& 0.52\% & 41.152	& 1.81\%	& 2193.391	& 96.71\%	& 2268.1 \\ \hline
DCMLDA 1324816 words & 22.658 & 0.65\%	& 10.52 & 0.30\% &	20.923	& 0.60\% & 3448.699	& 98.46\% &	3502.8 \\ \hline
DCMLDA 2596155 words & 22.658 & 0.28\% & 11.55 & 0.14\%	& 39.549 & 0.48\%	& 8153.969 & 99.10\%	& 8227.726 \\ \hline

\end{tabular}
\end{table*}

\subsection{Scaling-Up}

\begin{figure}[h]\centering
	\includegraphics[width=0.35\textwidth]{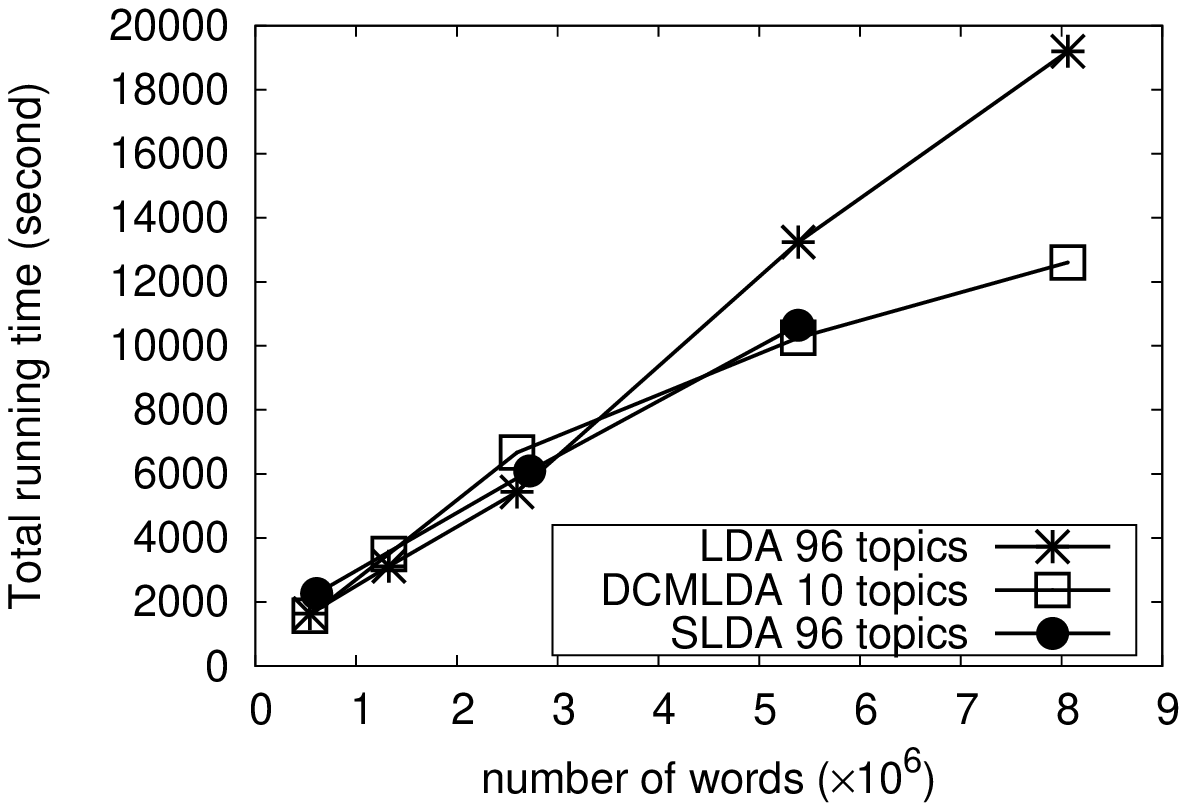}
	\caption{Scaling-up}
	\label{fig:scale-up}
\end{figure}

\figref{fig:scale-up} shows the total running time of LDA, SLDA, and DCMLDA on InferSpark
by scaling the data size (in words).
InferSpark scales well with the data size.
DCMLDA exhibits even super-linear scale-up. This is because as the data size goes up, 
the probability of selecting larger documents goes up. Consequently,
the growth in the total number of random variables is less than proportional, which gives rise
to the super-linearity.

\subsection{Scaling-Out}

\begin{figure}[h]
	\centering
	\includegraphics[width=0.35\textwidth]{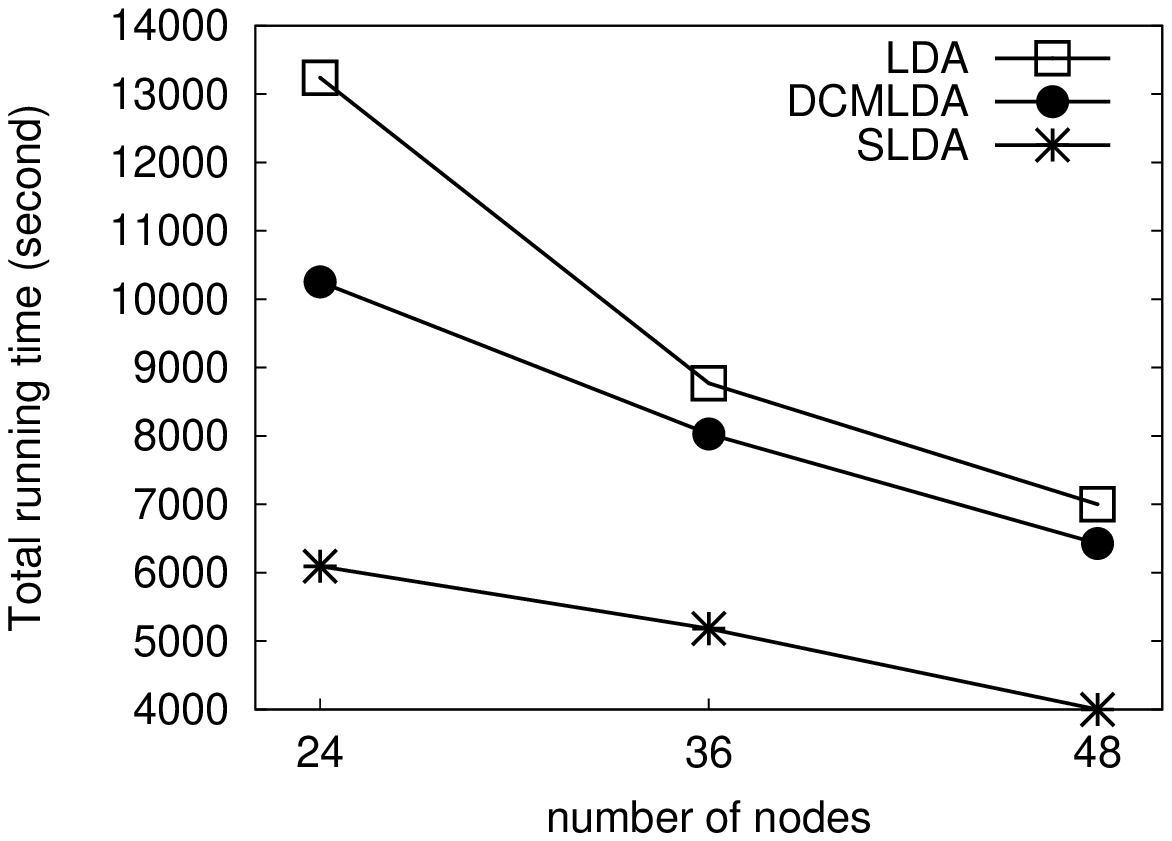}
	\caption{Scaling-out}
	\label{fig:scale-out}
\end{figure}

\figref{fig:scale-out} shows the total running time of LDA on InferSpark in
different cluster sizes. For each model, we use fixed size of dataset.  DCMLDA
and LDA both use the 2\% Wikipedia dataset. SLDA uses the 50\% amazon dataset.
We observe that InferSpark can achieve linear scale-out. 

\subsection{Partitioning Strategy}

\begin{figure}[h]\centering
	\includegraphics[width=0.35\textwidth]{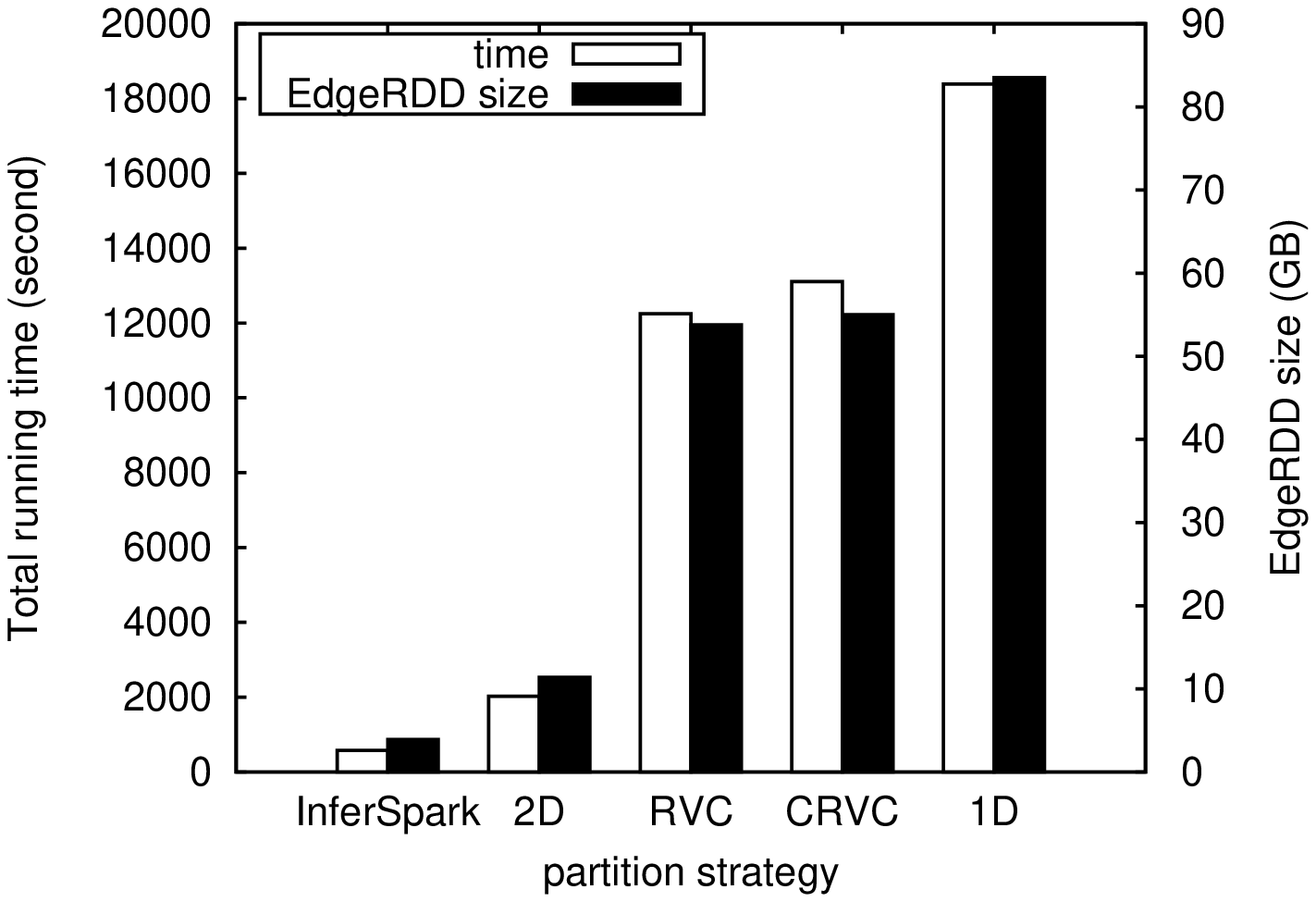}
	\caption{Comparison of Different Partition Strategies}
	\label{fig:exp_partition_strategy}
\end{figure}

\figref{fig:exp_partition_strategy} shows the running time of LDA(0.2\% Wikipedia dataset, 96 topics) on InferSpark
using our partitioning strategy  and 
GraphX partitioning strategies: 
EdgePartition2D  (2D)
RandomVertexCut (RVC),
CanonicalRandomVertexCut (CRVC), and
EdgePartition1D (1D).
We observe that the running time is propotional to the size of EdgeRDD.
Our partition strategy yields the best performance for running VMP on the
message passing graphs.  Our analysis shows that RVC and CRVC should have the
same results. The slight difference in the figure is caused by the randomness
of different hash functions.

%\begin{figure*}[!ht]
%	\tiny
%\begin{tabular}{|*{6}{l|}}
%	\hline
%	Experiement & Compilation (s) & Graph Building (s) & First Iteration (s) & Non-checkpointing iteration (s) & checkpointing iteration (s) \\\hline
%	LDA 2596155 words & 9.2 & 139.6 & 102.4 & 91.4 & 236.6 \\\hline
%	LDA 5386900 words & 9.1 & 302.8 & 309.8 & 226.2 & 555.2 \\\hline
%	LDA 8062163 words & 11.8 & 415.2 & 330.4 & 301.6 & 1050.4 \\\hline
%	SLDA 349569 words & 9.8 & 26.8 & 24.9 & 20.9 & 50.8 \\\hline
%	SLDA 607430 words & 10.4 & 41.8 & 42.8 & 38.6 & 94.5 \\\hline
%	DCMLDA 282660 words 96 topics & 11.0 & 29.9 & 163.1 & 141.4 & 336.0 \\\hline
%	DCMLDA 2596155 words 10 topics & 10.6 & 40.0 & 251.5 & 179.6 & 381.4 \\\hline
%\end{tabular}
%\caption{Time Breakdown (\%) of InferSpark (a) BN Construction; (b) Code Generation; (c) GraphX Inference}
%\label{fig:exp_times}
%\end{figure*}

%\begin{figure}[h]
%	\includegraphics[width=0.45\textwidth]{figs/exp_lda.eps}
%	\caption{Comprison of Average Iteration Time of LDA}
%	\label{fig:exp_lda}
%\end{figure}
%
%\begin{figure}[h]
%	\includegraphics[width=0.45\textwidth]{figs/exp_slda.eps}
%	\caption{Comprison of Average Iteration Time of Sentence-LDA}
%	\label{fig:exp_slda}
%\end{figure}
%
%\begin{figure}[h]
%	\includegraphics[width=0.45\textwidth]{figs/exp_dcmlda.eps}
%	\caption{Comparison of Average Iteration Time of Dirichlet Compound Multinomial LDA}
%	\label{fig:exp_dcmlda}
%\end{figure}

%\begin{figure}[h]
%	\includegraphics[width=0.45\textwidth]{figs/exp_slda_datasize.eps}
%	\caption{Inferspark Average Iteration Time vs Datasize of SLDA}
%	\label{fig:exp_slda_datasize}
%\end{figure}

%!TEX root = paper.tex

\section{Related Work}
\label{sec:related}

To the best of our knowledge, InferSpark is the only framework that can efficiently carry out statistical inference through probabilistic programming on a distributed in-memory computing platform.
MLlib, Mahout \cite{mahout}, and MADLib \cite{madlib} are machine learning \emph{libraries} on top of distributed computing platforms and relational engines.
All of them provide many standard machine learning models such as LDA and SVM.
However, when a domain user, say, a machine learning researcher, 
is devising and testing her customized models with her big data, 
those libraries cannot help.
MLBase \cite{mlbase} is related project that shares a different vision with us.
MLBase is a suite of {\em machine learning} algorithms and provides 
a declarative language for users to specify machine learning tasks.
Internally, it borrows the concept of query optimizer in traditional databases 
and has an optimizer that selects the best set of machine learning 
algorithms (e.g., SVM, Adaboost) for a specific task.
InferSpark, on the other hand, goes for a programming language approach, 
which extends Scala with the emerging probabilistic programming constructs,
and carries out {\em statistical inference} at scale. 
MLI \cite{mli} is an API on top of MLBase (and Spark) to ease the development of various distributed machine learning algorithms (e.g., SGD).
In the same vein as MLI, SystemML \cite{systemml} provides R-like language to ease the development of various distributed machine learning algorithms as well.
%!!SystemML [17] pro- poses an R-like language and shows how it can be optimized and compiled down to MapReduce. However, SystemML tries to support ML experts to develop efficient distributed algorithms and does not aim at simplifying the use of ML, for example, by automatically tuning the training step. Still, the the ideas of SystemML are compelling and we might leverage them as part of our physical plan optimization.
In \cite{Wang:2011} the authors present techniques to optimize inference algorithms in a probabilistic DBMS.

%% TODO TuPAQ and KeystoneML

There are a number of probabilistic programming frameworks other than Infer.NET \cite{InferNET14}.
For example, 
Church \cite{GMR+08} is a probabilistic programming language based on the functional programming language Scheme.
Church programs are interpreted rather than compiled.
Random draws from a basic distribution and queries about the execution trace are two additional type of expressions. A Church expression defines a generative model. Queries of a Church expression can be conditioned on any valid church expressions. Nested queries and recursive functions are also supported by Church. Church supports stochastic-memoizer which can be used to express nonparametric models. 
%Thus a wide range of probabilistic models can be concisely expressed in Church. %Inference in Church is implemented by rejection sampling and Metropolis-Hastings sampling algorithms. 
Despite the expressive power of Church, it cannot scale for large dataset and models. 
%Compared with InferSpark on the programming language level, 
%InferSpark programs are complied whereas Church programs are interpreted, where
%Our approach differs from Church in that our program is compiled rather than interpreted. Compiled program is usually more efficient than interpreted ones. We can also leverage the distributed parallel processing to scale up to large dataset and models.
Figaro is a probabilistic programming language implemented as a library in Scala \cite{Figaro}.
It is similar to Infer .NET in the way of defining models and performing inferences but put more emphasis to object-orientation. Models are defined by composing instances of Model classes defined in the Figaro library.
Infer.NET is a probabilistic programming framework in C\# for Bayesian Inference. A rich set of variables are available for model definition. Models are converted to a factor graph on which efficient built-in inference algorithms can be applied. %The algorithms include expectation propagation, variational message passing, block Gibbs sampling. Infer .NET addresses the scalability by shared variables, as discussed in Section \ref{sec:intro}.
Infer.NET is the best optimized probabilistic programming frameworks so far.
Unfortunately, all existing probabilistic programming frameworks including Infer.NET cannot scale out on to a distributed platform.

%%%%%%%%%%%%%%%%

%!TEX root = paper.tex

\section{Conclusion}
\label{sec:conclusion}

%In this paper, we propose a new probabilistic programming framework built on Spark. We believe the framework can make it easier to develop probabilistic models and apply Bayesian inference on large distributed dataset. It will enable domain experts to create more complex and powerful probabilistic models.
This paper presents InferSpark, 
a probabilistic programming framework  on Spark.
 Probabilistic programming is an emerging paradigm that allows statistician and domain users to succinctly express a statistical model 
 within a host programming language and transfers the burden of
implementing the inference algorithm from the user to the compilers and
runtime systems.
InferSpark, to our best knowledge, is the first probabilistic programming  framework
that builts on top of a distributed computing platform.
Our empirical evaluation shows that InferSpark can successfully express 
some known Bayesian models in a very succinct manner
and can carry out distributed inference at scale.
InferSpark will open-source.  
The plan is to invite the community to extend InferSpark to support other types of 
statistical models (e.g., Markov networks) and to support more kinds of inference techniques 
(e.g., MCMC).

\section{Future Directions}
\label{sec:future}

Our prototype InferSpark system only implements the variational message
passing inference algorithm for certain exponential-conjugate family Bayesian
networks (namely mixtures of Categorical distributions with Dirichlet priors).
In our future work, we plan to include support for other common types of
Bayesian networks (e.g. those with continuous random variables or arbitrary
priors). The VMP algorithm may be no longer applicable to these Bayesian
networks because they may have non-conjugate priors or distributions out of
exponential family. In order to handle wider classes of graphical models, we
also plan to incorporate other inference algorithms (e.g. Belief propagation,
Gibbs Sampling) into our system, which could be quite challenging because we
have to 1) deal with arbitrary models 2) adapt the algorithm to distributed
computing framework.

Another interesting future direction is to allow implementation of customized
inference algorithms as plugins to the InferSpark compiler. To make the
development of customized inference algorithms in InferSpark easier than
directly writing them in a distributed computing framework, we plan to 1)
revise the semantics of the Inferspark model definition language and expose a
clean Bayesian network representation 2) provide a set of framework-independent
operators for implementing the inference algorithms 3) investigate how to
optimize the operators on Spark.

\bibliographystyle{abbrv}
\bibliography{paper}

\appendix
%\input{syntax}
%!TEX root = paper.tex
\section{SLDA and DCMLDA in InferSpark}\label{models}

\begin{figure}[h]
\begin{lstlisting}
@Model
class SLDA(K: Long, V: Long, alpha: Double, beta: Double) {
  val phi = for (i <- 0L until K) yield Dirichlet(beta, V)
  val theta = for (i <- ?) yield Dirichlet(alpha, K)
  val z = theta.map{theta => for (i <- ?) yield Categorical(theta)}
  val x = z.map(_.map(z => ?.map(_ => Categorical(phi(z)))))
}
\end{lstlisting}
	\caption{SLDA Model in InferSpark}
\end{figure}

\begin{figure}[h]
\begin{lstlisting}
@Model 
class DCMLDA(K: Long, V: Long, alpha: Double, beta: Double) {
    val doc = for (i <- ?) yield {
        val phi = (0L until K).map(_ => Dirichlet(beta, V))
        val theta = Dirichlet(alpha, K)
        val z = ?.map(_ => Categorical(theta))
        val x = z.map(z => Categorical(phi(z)))
    }
}
\end{lstlisting}
	\caption{DCMLDA Model in InferSpark}
\end{figure}

\end{document}